\documentclass[a4paper,fleqn]{cas-sc}

\usepackage[numbers]{natbib}

\def\tsc#1{\csdef{#1}{\textsc{\lowercase{#1}}\xspace}}
\tsc{WGM}
\tsc{QE}
\tsc{EP}
\tsc{PMS}
\tsc{BEC}
\tsc{DE}

\usepackage{amsthm}
\usepackage{amsfonts}
\usepackage{amssymb}
\usepackage{mathrsfs}
\usepackage{mathtools}
\usepackage[english]{babel}
\usepackage{float}
\usepackage{textcomp}

\usepackage{epstopdf}
\usepackage{amsmath}
\usepackage{xcolor}
\usepackage{url}
\usepackage{multirow}
\usepackage{graphicx}
\newcommand{\expec}[1]{\mathbb{E}\left[#1\right]}
\DeclareMathOperator*{\argmin}{arg\,min}

\usepackage{bm}
\usepackage{balance}
\usepackage{microtype}
\usepackage[caption=false,font=footnotesize]{subfig}
\usepackage{comment}

\allowdisplaybreaks

\newtheorem{theorem}{Theorem}
\makeatletter
\newcommand{\settheoremtag}[1]{
  \let\oldthetheorem\thetheorem
  \renewcommand{\thetheorem}{#1}
  \g@addto@macro\endtheorem{
    \addtocounter{theorem}{-1}
    \global\let\thetheorem\oldthetheorem}
  }
\makeatother

\newcommand{\suml}{\sum_{\ell=1}^{L}}
\newcommand{\sump}{\sum_{p=1}^{P}}
\newcommand{\sumk}{\sum_{k=1}^{K}}
\newcommand{\sumq}{\sum_{q=1}^{Q}}

\newcommand{\bsym}{\boldsymbol}

\newcommand{\en}[1]{\left[#1\right]}
\newcommand{\alpharl}{[\bsym{\alpha}_r]_\ell}
\newcommand{\taurl}{[\bsym{\overline{\tau}}_r]_\ell}
\newcommand{\nurl}{[\bsym{\overline{\nu}}_r]_\ell}
\newcommand{\alphacq}{[\bsym{\alpha}_c]_q}
\newcommand{\taucq}{[\bsym{\overline{\tau}}_c]_q}
\newcommand{\nucq}{[\bsym{\overline{\nu}}_c]_q}
\DeclareMathOperator*{\minimize}{minimize }
\DeclareMathOperator*{\maximize}{maximize }

\newtheorem{definition}[theorem]{Definition}

\newtheorem{lemma}[theorem]{Lemma}

\newtheorem{proposition}[theorem]{Proposition}
\newtheorem{remark}[theorem]{Remark}

\begin{document}
\shorttitle{Multi-Antenna Dual-Blind Deconvolution}


\title [mode = title]{Multi-Antenna Dual-Blind Deconvolution for Joint Radar-Communications via SoMAN Minimization}    


\tnotetext[1]{This research was sponsored by the Army Research Office/Laboratory under Grant Number W911NF-21-1-0099, and the VIE project entitled ``Dual blind deconvolution for joint radar-communications processing''. This paper was supported by the Vicerrectoría de Investigacion y Extensión UIS under the research projects 3735 and 3924.  K. V. M. acknowledges support from the National Academies of Sciences, Engineering, and Medicine via the Army Research Laboratory Harry Diamond Distinguished Fellowship. The following co-authors are a member of EURASIP: Kumar Vijay Mishra.}

%
\author[label1]{Roman Jacome}[orcid=0000-0002-3621-0275]
\author[label1]{Edwin Vargas}[orcid=0000-0002-7979-9497]
\author[label2]{Kumar Vijay Mishra}[orcid=0000-0002-5386-609X]
\author[label2]{Brian M. Sadler}[orcid=0000-0002-9564-3812]
\author[label1]{Henry Arguello}[orcid=0000-0002-2202-253X]

\affiliation[label1]{organization={Universidad Industrial de Santander},
            city={Bucaramanga},
            postcode={680002}, 
            country={Colombia}}
\affiliation[label2]{organization={United States DEVCOM Army Research Laboratory},
            city={Adelphi},
            postcode={MD 20783}, 
            country={United States}}

\maketitle

\begin{abstract}
In joint radar-communications (JRC) applications such as secure military receivers, often the radar and communications signals are overlaid in the received signal. In these passive listening outposts, the signals and channels of both radar and communications are unknown to the receiver. The ill-posed problem of recovering all signal and channel parameters from the overlaid signal is termed as \textit{dual-blind deconvolution} (DBD). In this work, we investigate DBD for a multi-antenna receiver. We model the radar and communications channels with a few (sparse) \textit{continuous-valued} parameters such as time delays, Doppler velocities, and directions-of-arrival (DoAs). To solve this highly ill-posed DBD, we propose to minimize the sum of multivariate atomic norms (SoMAN) that depend on unknown parameters. To this end, we devise an exact semidefinite program using theories of positive hyperoctant trigonometric polynomials (PhTP). Our theoretical analyses show that the minimum number of samples and antennas required for perfect recovery is logarithmically dependent on the maximum of the number of radar targets and communications paths rather than their sum. We show that our approach is easily generalized to include several practical issues such as gain/phase errors and additive noise. Numerical experiments show the exact parameter recovery for different JRC scenarios.
\end{abstract}

\section{Introduction}
\label{sec:intro}
The existing spectrum allocation between radar systems and wireless communications is designed by various regulatory bodies with the aim of avoiding user interference at all times \cite{griffiths2015radar}. In view of the increasing demand for throughput and high sensing resolution, this conservative approach to spectrum access is not optimal and often results in reduced performance. This has led to the development of spectrum-sharing joint radar-communications (JRC) systems, which also have the advantages of low cost, compact size, and less power consumption \cite{mishra2019toward,wu2022resource,liu2020co}. In many JRC applications, a common waveform is transmitted and a joint receiver decodes the communications symbols and estimates the radar target parameters \cite{elbir2021terahertz,dokhanchi2019mmwave}. In such cases, conventionally, radar transmits waveform (communications channel) is known (estimate \textit{a priori}) at the receiver, whose goal is to estimate the unknown target parameters (transmitted communications messages). In this paper, we focus on the case when transmitting signals as well as channels of both radar and communications are unknown to a common receiver.

The aforementioned scenario is common in military listening passive devices \cite{vargas2022joint}, where deciphering overlaid radar and communications information is required with minimum prior knowledge. In a passive \cite{sedighi2021localization} or multistatic \cite{dokhanchi2019mmwave} radar, the receiver is generally not aware of the transmit waveform. In millimeter-wave \cite{mishra2019toward} or terahertz-band \cite{elbir2021terahertz} wireless communications, the channel coherence times are very short and channel estimates quickly become outdated. Additionally, when the received communications signal is overlaid with radar target echoes, it is not possible to accurately estimate the communications channel. In previous studies \cite{vargas2023dual,vargas2022joint,monsalve2023beurling}, the extraction of all four of these quantities, i.e., radar and communications channels and signals, was modeled as an ill-posed \textit{dual-blind deconvolution} (DBD) problem, wherein the observation is a sum of two convolutions and the goal is to estimate all four convolvands. This formulation is related to the long-standing signal processing problem of (single-)blind deconvolution (BD) that is encountered in many applications \cite{jefferies1993restoration,ayers1988iterative,abed1997blind}. The non-blind \textit{dual deconvolution} has been reported for overlaid receivers in \cite{farshchian2016dual}.
 
 In this paper, contrary to previous studies \cite{vargas2023dual,vargas2022joint,monsalve2023beurling,jacome2023factor} that employed a single antenna receiver, we focus on a more challenging DBD scenario that requires a multi-antenna joint receiver to also estimate the direction of arrival (DoA) of radar targets and communications signals besides their conventional delay-Doppler channels. In particular, we consider the radar transmit signal to be a train of pulses and the communications waveform follows orthogonal frequency division multiplexing (OFDM). The channel parameters - delay, Doppler, and DoA - are \textit{continuous-valued} and sparse. Prior state-of-the-art studies \cite{tang2013compressed,xu2014precise,mishra2015spectral,chi2016guaranteed} show that continuous-valued parameters in such ill-posed problems are precisely recovered using the framework of \textit{atomic norm minimization} (ANM). In particular, \cite{vargas2023dual} addressed a single-antenna DBD by minimizing a sum of multivariate atomic norms (SoMAN).

 However, traditional SoMAN minimization in DBD does not trivially extend to the multi-antenna scenario because its recovery guarantees are coupled, i.e., the recovery of radar parameters from the overlaid signal is not independent of communications parameters and vice versa \cite{vargas2023dual}. In multi-antenna DBD (M-DBD), this coupling must be additionally resolved over the DoA domain. 
 Our work has close connections with the rich heritage of research on super-resolved (non-blind) DoA estimation \cite{yang2018sparse,heckel2016superMIMO}, including using ANM \cite{tang2019grid,wu2022gridless}. A related problem is the multichannel blind deconvolution (MCBD) but it considers convolution of a single transmit signal with multiple propagation channels 
 \cite{lee2018fast}. The DBD is also connected to conventional blind demixing (BdM), where the observation is a mixture of multiple BDs \cite{ling2017blind}. However, unlike DBD with different radar and communications signal structures, all convolutions in BdM have an \textit{identical} form. We refer the interested reader to \cite[Table I]{vargas2023dual} for detailed differences between DBD and similar ill-posed problems.

Preliminary results of our work appeared in \cite{jacome2022multid}, where the noiseless case of three-dimensional (3-D) SoMAN formulation was derived. In this paper, we apply our approach to estimate the communications messages and radar waveforms, formulate 3-D SoMAN in the presence of noise, provide detailed theoretical guarantees, incorporate steering vector errors in both gain and phase, address n-tuple M-DBD, and include additional numerical validations. Our main contributions are:\\ 
\textbf{1) M-DBD with structured unknown continuous-valued parameters.} We exploit the sparsity of both radar and communications channels to formulate the recovery of unknown continuous-valued channel/signal parameters as a 3-D DBD problem. Following the approaches in \cite{vargas2023dual,yang2016super}, we represent the unknown transmit radar signal (a periodic waveform) and communications messages in a low-dimensional subspace spanned by the columns of a known representation basis. This representation allows including the special structure of radar and communications signals in our M-DBD formulation.\\
\textbf{2) 3-D SoMAN-based recovery.} We formulate our problem as the minimization of the sum of two tri-variate atomic norms. However, the primal SoMAN problem does not directly yield a semidefinite program (SDP). We, therefore, turn to the dual problem and derive the SDP using the theories of positive hyperoctant trigonometric polynomials (PhTP) \cite{dumitrescu2017positive}. In the non-blind case, this approach has been previously employed for high-dimensional super-resolution (SR) \cite{xu2014precise} and bivariate radar parameter estimation \cite{heckel2016super}. We demonstrate our approach through extensive numerical experiments.\\
\textbf{3) Strong theoretical guarantees.} We provide a theoretical guarantee on the number of samples, number of pulses/messages, and number of received antennas for exact recovery with high probability. This result is based on constructing PhTPs that achieve the conditions of the dual certificate. We show that the minimum number of overall samples for perfect recovery scale logarithmically with the maximum of the radar targets and communications paths rather than their sum.\\
\textbf{4) Practical issues.} We consider practical issues to generalize our approach. In the presence of noise, our formulation adds a regularization term to the dual problem. We further show that our method is robust to both gain and phase errors in the steering vector \cite{liao2013recursive} and derive the optimality of the corresponding regularization parameters. In the non-blind scenarios, these errors have been tackled through techniques such as eigendecomposition of the measurement covariance matrix \cite{liu2011eigenstructure}, Hadamard product-based estimation \cite{cao2013hadamard}, and regularized ANM \cite{chen2020new}.

The rest of the paper is organized as follows. In the next section, we describe the signal model for the multi-antenna overlaid JRC receiver. We devise the exact SoMAN SDP for 3-D DBD recovery in Section~\ref{sec:formulation} along with the procedure to recover radar and communications waveform and theoretical recovery guarantees. The practical scenarios of noise and antenna errors are considered in Section \ref{sec:errors}, wherein we also provide optimal regularization parameters. The proof of our main result is detailed in Section ~\ref{sec:proof}. We validate our model and methods through extensive numerical experiments in Section~\ref{sec:experiments} and conclude in Section~\ref{sec:summ}. 
\begin{table}[!t]

    \centering
    \caption{Glossary of notations}
    \begin{tabular}{|c||l|}\hline
         \textbf{Symbol}&\textbf{Definition} \\\hline 
          $\mathcal{A}_r$ ($\mathcal{A}_c$)& Atomic set of radar (communications) signal\\
                   $\alpharl$ ($\alphacq$)& Propagation loss parameter of radar (communications) channel\\
                 $B$& Signal bandwidth \\

          $\mathcal{B}_r$ ($\mathcal{B}_c$)& Radar (communications) sensing linear operator\\
                $[\bsym{\beta}_r]_\ell$ ($[\bsym{\beta}_c]_q$)& DoA parameter of radar (communications) channel\\
                 $f_k$ & subcarrier frequencies \\
               $\mathbf{f}_r(\mathbf{r})$ ($\mathbf{f}_c(\mathbf{c})$)& Positive hyperoctant trigonometric polynomial for radar (communciations)\\
                 $[\mathbf{g}_p]_k$& Transmitted symbol\\
                         
                $\mathbf{G}_{\widetilde{m}}$ ($\mathbf{A}_{\widetilde{m}}$)& Sensing matrix of radar (communications) \\
                 $h_r(t)$ ($h_c(t)$)& Radar (communications) channel\\
                         $J$& Low dimensional subspace size\\
         $K$ & Number of subcarriers \\
         $L$& Number of radar targets\\
$\lambda$ & Operating wavelength\\
         $\mu$& Incoherence parameter\\

$\mu_r$ $(\mu_c)$& Regularization parameter for radar (communications)\\
                 $\mathbf{n}$& Vector gain error\\
                   $N$& Number of frequency samples\\
         $N_R$& Number of received antennas\\
               $\nurl$($\nucq$)& Doppler parameter for radar (communications) channel\\
         $\mathbf{q}$& Dual variable\\
         
         $Q$& Number of communications propagation paths\\
         
         $\mathbf{Q}$& Gram matrix\\
  $\mathbf{\widehat{Q}}_r$ ($\mathbf{\widehat{Q}}_c$)& Coefficient of the 3-D radar (communications) trigonometric polynomials\\
           $\mathbf{r}_\ell$ ($\mathbf{c}_q$)& Vectors containing radar (communications) channel parameters\\
           $S(t)$ & Transmitted radar waveform \\
         
                  $T$& PRI or symbol duration \\
 $\mathbf{T}$ ($\mathbf{D}$)& Low dimensional basis of radar (communications) signal\\
  $\taurl$ ($\taucq$)& Delay parameter for radar (communications) channels\\
         $\mathbf{u}$ ($\mathbf{v}$)& Low dimensional coefficient vector of radar (communications) signals\\
           $\bsym{\phi}$& Vector phase error\\
         $x_r(t)$ ($x_c(t)$)& Transmitted radar (communications) signal\\
        
         $\mathbf{X}_r$ ($\mathbf{X}_c$)& Lifted matrix of the unknown variables of radar (communications) signals\\
                $[\mathbf{y}_{\widetilde{m}}]$ & Discrete samples of received overlaid signal\\
\hline
    \end{tabular}
    \label{tab:notations}
\end{table}

\begin{table}[!t]

    \centering
    \caption{Glossary of acronyms}
    \begin{tabular}{|c||l|}\hline
\textbf{Acronym}& \textbf{Definition}\\\hline
         ANM& Atomic norm minimization\\
          BD& Blind deconvolution\\
          BRL&Bounded real lemma\\
          CPI&coherent processing interval\\
          CTFT& Continuous-time Fourier transform\\
          DBD& Dual blind deconvolution\\
         DoA& Direction of arrival\\
         JRC& Joint radar-communications\\ 
         KKT&Karush–Kuhn–Tucker\\      
         LMI&Linear matrix inequality\\

         M-DBD&Multi-antenna dual blind deconvolution\\
          OFDM& orthogonal frequency division
multiplexing\\    
         PhTP& Positive hyperoctant trigonometric polynomials \\

 PRI& Pulse repetition interval\\
         SoMAN& Sum of multivariate atomic norms\\
         SDP& Semidefinite program\\
         ULA&Uniform linear array\\\hline 
    \end{tabular}
    \label{tab:acronyms}
\end{table}
Throughout this paper, we reserve boldface lowercase, boldface uppercase, and calligraphic letters for vectors, matrices, and index sets, respectively. The notation $\en{\mathbf{x}}_i$ indicates the $i$-th entry of the vector $\mathbf{x}$ and $\en{\mathbf{X}}_i$ the $i$-th row of $\mathbf{X}$. We denote the transpose, conjugate, and Hermitian by $(\cdot)^T$, $(\cdot)^*$, and $(\cdot)^H$, respectively. The identity matrix of size $N\times N$ is $\mathbf{I}_N$. $||\cdot||_p$ is the $\ell_p$ norm. The notation $\text{Tr}\left\lbrace \cdot \right\rbrace $ is the trace of the matrix, $|\cdot|$ is the cardinality of a set, $\operatorname{supp}(\dot)$ is the support set of its argument, $\mathbb{E}\left[ \cdot \right]$ is the statistical expectation function, and $\mathbb{P}$ denotes the probability. The functions $\text{max}$ and $\text{min}$ output their arguments' maximum and minimum values, respectively. The sign function is defined as $\operatorname{sign}(c) = \frac{c}{|c|}$. The function $\text{diag}(\cdot)$ outputs a diagonal matrix with the input vector along its main diagonal. The block diagonal matrix with diagonal matrix elements $\mathbf{X}_1$, $\dots$, $\mathbf{X}_P$ is  $\mathbf{X} = \operatorname{blockdiag}[\mathbf{X}_1,\dots,\mathbf{X}_P]$. Table \ref{tab:notations} and \ref{tab:acronyms} summarizes the mathematical symbols and acronyms employed in the paper respectively.

\section{System Model}\label{sec:system}
\begin{figure}[t]
    \centering
    \includegraphics[width=0.56\linewidth]{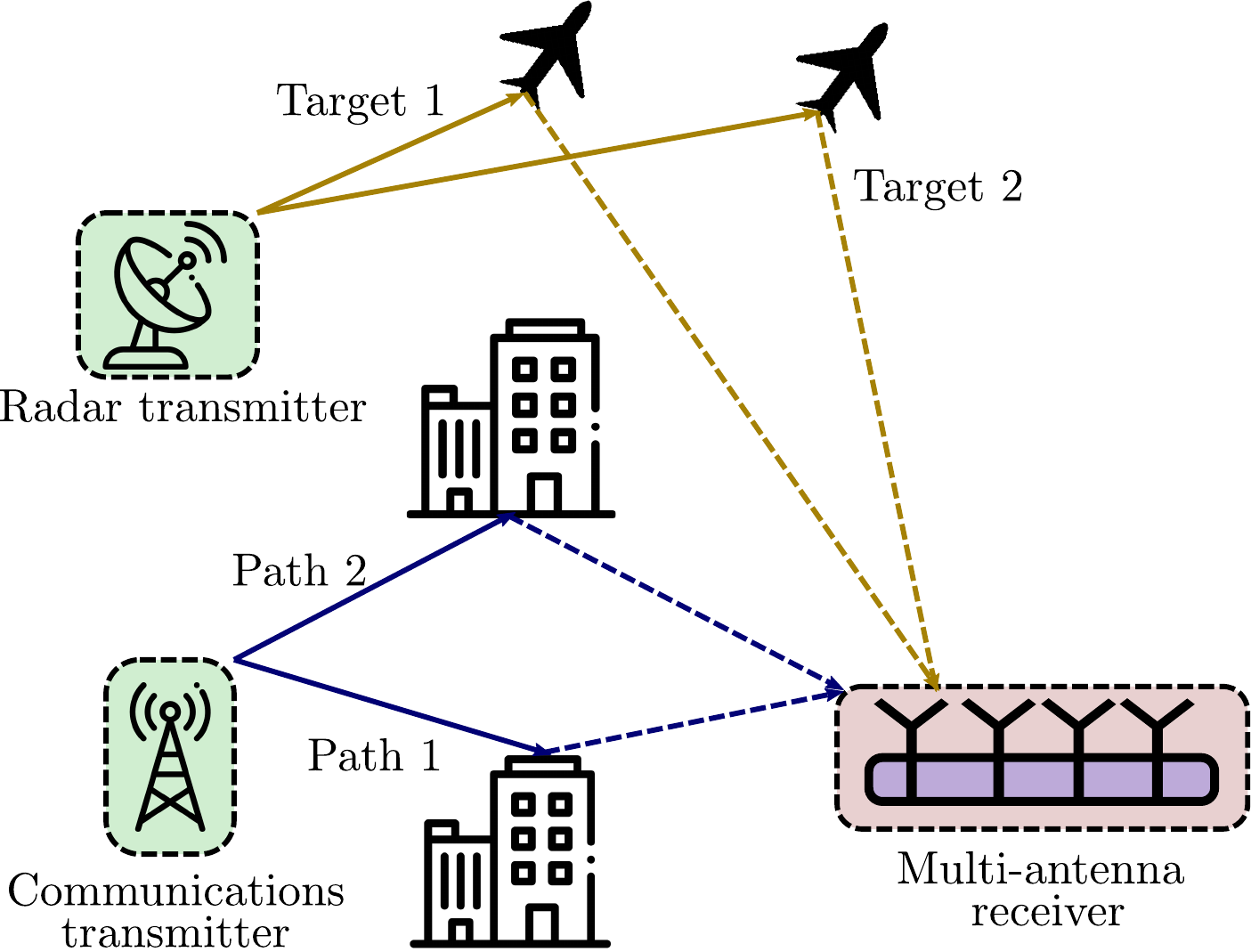}
    \caption{The JRC ULA receive antenna admits a superposition of independently transmitted radar and communications signals that are reflected off scatterers along their respective paths.}
    \label{fig:system}
\end{figure}

We consider an overlaid JRC receiver that employs a uniform linear array (ULA) (Fig.~\ref{fig:system}) with $N_R$ antennas and an inter-element spacing of $d=\lambda/2$, where $\lambda=c/f_c$ is the wavelength, $f_c$ is the carrier frequency, and $c=3\times 10^8$ m/s is the speed of light. It receives a superposition of unknown radar and communications signals that were transmitted independently of each other and traversed to the receiver through corresponding unknown channels. 

\subsection{Transmit signals}
Consider a radar that transmits a train of $P_r$ pulses $s(t)$ at a pulse repetition interval (PRI) $T_r$ as
\begin{equation}
    x_r (t) = \sum_{p=0}^{{P_r}-1} s(t-pT_r),\;\; 0<t<({P_r}-1)T_r,
\end{equation}
The pulse $s(t)$ is a time-limited baseband function, whose continuous-time Fourier transform (CTFT) is $S(f)=\int_{-\infty}^{\infty} s(t) e^{-j 2\pi f t} \mathrm{d}t$. In general, most of the radar signal's energy lies within the spectrum $[-B/2, B/2]$, where $B$ denotes the effective signal bandwidth, i.e., $s(t) \approx \int_{-B/2}^{B/2} S(f) e^{j 2\pi f t} \mathrm{d}f$. The entire duration of $P_r$ pulses is known as the coherent processing interval (CPI). 

The communications transmit signal denoted {by} $x_c(t)$ is the conventional orthogonal frequency-division multiplexing (OFDM) signals with $K$ equi-bandwidth sub-carriers. These subcarriers together occupy a bandwidth of $B$ with a symbol duration $T_c$ and frequency separation  $\Delta f$ \cite{cimini1985analysis}. The communications system transmits $P_c$ messages, where the $p$-th transmitted message is $g_p(t) = \sumk [\mathbf{g}_p]_k e^{\mathrm{j}2\pi k\Delta f t}$, where $[\mathbf{g}_p]_k$ is $p$-th complex symbol modulated onto the $k$-th sub-carrier frequency $f_k=k\Delta f$. The communications transmit signal is 
\begin{align}
x_c(t) = \sum_{p = 0}^{P_c-1} g_p(t-pT_c),\; 0\leq t \leq P_c T_c. \label{comm_transmitted}
\end{align}

\vspace{-10pt}
\subsection{Channels}
The radar channel or \textit{target scene} comprises $L$ non-fluctuating point-targets, according to the Swerling-I target model \cite{skolnik2008radar}. 
The unknown target parameter vectors are ${\bsym{\alpha}}_r \in \mathbb{C}^L$, ${\overline{\bsym{\tau}}}_r \in \mathbb{R}^L$, ${\overline{\bsym{\nu}}}_r\in \mathbb{R}^L$, ${{\bsym{\beta}}}_r\in \mathbb{R}^L$ and where the $\ell$-th target is characterized by: time delay $\taurl$, which is linearly proportional to the target's range \textit{i.e.} $\taurl = 2d_\ell/c$ where $c$ is the speed of light and $d_\ell$ is the target range; Doppler frequency $\nurl$, proportional to the target's radial velocity i.e., $ {\nurl} = 4\pi f_c v_\ell/c $ where $v_\ell$ is the radial velocity; DoA $[\bsym{\beta}_r]_\ell = \sin(\theta_k)/2$ where $\theta_k$ is the relative angle to the received array and complex amplitude $[\bsym{\alpha}_r]_\ell$ that models the path loss and reflectivity of the $\ell$-th target. The target locations are defined with respect to the polar coordinate system of the radar and their range and Doppler are assumed to lie in the unambiguous time-frequency region, i.e., the time delays ($\tau$) are no longer than the PRI, and Doppler frequencies ($\nu$) are up to the PRF. The radar channel impulse response is  
\begin{align}
    h_r(\nu, \tau, \beta) = \suml [\bsym{\alpha}_r]_\ell \delta(\nu-\nurl) \delta(\tau-\taurl)\delta(\beta-[\bsym{\beta}_r]_\ell).\label{eq:rad_channel1}
\end{align}
The delay-Doppler-DoA representation of the channel is obtained by taking the Fourier transform along the time and spatial axes  of the time-delay representation as 
\begin{align}
    h_r(t, \tau, x) = \suml [\bsym{\alpha}_r]_\ell \delta(\tau-[{\overline{\bsym{\tau}}_r} ]_\ell)e^{-\mathrm{j}2\pi \nurl t}e^{-\mathrm{j}2\pi x [\bsym{\beta}_r]_\ell}.\label{eq:radar_channel}
\end{align}
Considering the antenna array receiver, we discretize the spatial dimension given by the antenna separation, the channel at the location $x_m = \frac{2md}{\lambda}$ of the $m$-th antenna is \cite{johnson1992array,van2004detection,chen2020new}
\begin{align}
    h_r(t,  \tau, x_m) 
    &= \suml [\bsym{\alpha}_r]_\ell \delta(\tau-[{\overline{\bsym{\tau}}_r} ]_\ell)e^{-\mathrm{j}2\pi \nurl t}e^{-\mathrm{j}2\pi x_m [\bsym{\beta}_r]_\ell}\nonumber\\&= \suml [\bsym{\alpha}_r]_\ell \delta(\tau-[{\overline{\bsym{\tau}}_r} ]_\ell)e^{-\mathrm{j}2\pi \nurl t}e^{-\mathrm{j}2\pi \frac{2md}{\lambda} [\bsym{\beta}_r]_\ell}.
    \label{eq:radar_channel_2}
\end{align}
Concatenating the values of  $h_r(t, x_m, \tau) $, $m=1,\dots,N_R$ in the entries of the vector  $\mathbf{h}_r(t,\tau) \in \mathbb{C} ^{N_R}$, 
the radar channel for the $m$-th antenna is 
\begin{align}
    \en{\mathbf{h}_r(t,\tau)}_m = \suml [\bsym{\alpha}_r]_\ell \delta(\tau-[{\overline{\bsym{\tau}}_r} ]_\ell)e^{-\mathrm{j}2\pi \nurl t}e^{-\mathrm{j}2\pi m [\bsym{\beta}_r]_\ell}.\label{eq:radar_channel_final}
\end{align}
The communications channel comprises $Q$ propagation paths characterized by their attenuation coefficients, time-delay, Doppler velocity, and DoA, respectively, encapsulated in the parameter vectors $\bsym{\alpha}_c \in \mathbb{C}^Q$, $\overline{\bsym{\tau}}_c \in \mathbb{R}^Q$, $\overline{\bsym{\nu}}_c \in \mathbb{R}^Q$, and  $\bsym{\beta}_c \in\mathbb{R}^Q$. The delay-Doppler-DoA representation of the communications channel is
\begin{equation}
\label{eq:com_channel1}
    {h}_c(\nu,\tau,\beta) = \sumq \alphacq \delta(\nu-\nucq) \delta (\tau-\taucq)\delta(\beta-[\bsym{\beta}_c]_q).
\end{equation}
Computing the Fourier transform in the time and spatial domain  and considering the ULA receiver, we obtain the time-delay communications channel representation for the $m$-th antenna as
\begin{equation}
    \en{\mathbf{h}_c(t,\tau)}_m = \sumq \alphacq \delta (\tau-\taucq)e^{-\mathrm{j}2\pi\nucq t}e^{-\mathrm{j}2\pi m[\bsym{\beta}_c]_q}. \label{eq:comms_channel}
\end{equation}
\vspace{-10pt}
\subsection{Overlaid Radar-Communications Multi-Antenna Receiver}
The received signal $\mathbf{y}(t) = \big[ [\mathbf{y}(t)]_1, \dots, [\mathbf{y}(t)]_{N_R} \big]^T $ for all $N_R$ antennas is the superposition of radar and communications signals, where the signal at the $m$-th receive antenna is
\begin{align}
[\mathbf{y}(t)]_m= &\int\limits_{-\infty}^\infty[\mathbf{h}_r(t,\tau)]_mx_r(t-\tau)+[\mathbf{h}_c(t,\tau)]_mx_c(t-\tau) d\tau\nonumber\\&=[\mathbf{h}_r(t)]_m*x_r(t) + [\mathbf{h}_c(t)]_m*x_c(t),\;m=1,\dots,N_R.\label{eq:joint_receiver}
\end{align}
Substituting the expressions of the radar and communications channels from, respectively, \eqref{eq:radar_channel} and \eqref{eq:comms_channel} as well as the transmit signals $x_r(t)$ and $x_c(t)$ in \eqref{eq:joint_receiver}, we obtain
\begin{align}
    &[\mathbf{y}(t)]_m \nonumber\\
    &= \suml\sum_{p=1}^{P_r} \alpharl  s(t-pT_r - \taurl)e^{-\mathrm{j}2\pi \nurl t}e^{-\mathrm{j}2\pi m[\bsym{\beta}_r]_\ell} +\sumq\sum_{p=1}^{P_c} \alphacq g_p(t-pT_c-\taucq)e^{-\mathrm{j}2\pi \nucq t}e^{-\mathrm{j}2\pi m[\bsym{\beta}_c]_q}\nonumber\\&\approx  \suml\sum_{p=1}^{P_r} \alpharl  s(t-pT_r - \taurl)e^{-\mathrm{j}2\pi \nurl pT_r}e^{-\mathrm{j}2\pi m[\bsym{\beta}_r]_\ell} +\sumq\sum_{p=1}^{P_c} \alphacq g_p(t-pT_c-\taucq)e^{-\mathrm{j}2\pi \nucq pT_c}e^{-\mathrm{j}2\pi m[\bsym{\beta}_c]_q},\label{eq:ym_approx}
\end{align}
where the last approximation follows from $[\overline{\bsym{\nu}}_r]_\ell T_r\ll1$ and $[\overline{\bsym{\nu}}_c]_q T_c \ll 1$ \cite{zheng2017super}, which results in a constant phase rotation within one CPI.

For the sake of simplicity,\footnote{These assumptions, including the implication that both radar and communications transmissions are synchronized, are aimed to simplify the mathematical notations in the sequel. In the case of unequal durations and unsynchronized transmissions, we refer the reader to \cite[Section V]{vargas2023dual}, where we demonstrated the procedure to include the practical issue of unsynchronized transmission in the DBD problem.} the number of pulses equals the number of messages, i.e., $P_r=P_c = P$ and the PRI $T_r$ is the same as the message duration $T_c$ \textit{i.e.} $T_r = T_c = T$. Then, \eqref{eq:ym_approx} becomes
\begin{align}
    &[\mathbf{y}(t)]_m \nonumber\\
    &= \sump \left(\suml \alpharl  s(t-pT - \taurl)e^{-\mathrm{j}2\pi \nurl pT}e^{-\mathrm{j}2\pi m[\bsym{\beta}_r]_\ell}+ \sumq \alphacq g_p(t-pT-\taucq)e^{-\mathrm{j}2\pi \nucq pT}e^{-\mathrm{j}2\pi m[\bsym{\beta}_c]_q}\right)\nonumber\\&\triangleq \sum_{p=1}^{P-1}[\widetilde{\mathbf{y}}_p(t)]_m.
\end{align}
Re-write the measurements based on the shifted signals i.e. $[\mathbf{y}_p(t)]_m =[\widetilde{\mathbf{y}}_p(t+ pT)]_m$ that are time-aligned with $[\mathbf{y}_0(t)]_m$ where all $[\mathbf{y}_p(t)]_m$, for $p=1,\dots,P$, contain the same delay, Doppler, and DoA parameters. The continuous-time Fourier transform (CTFT) of $[\mathbf{y}_p(t)]_m$ is
\begin{align}
    &[\mathbf{\widehat{y}}_p(f)]_m  \nonumber\\&=\int\limits_{pT}^{pT+T} e^{-\mathrm{j}2\pi ft} \left(\suml \alpharl  s(t- \taurl)e^{-\mathrm{j}2\pi \nurl pT}e^{-\mathrm{j}2\pi m[\bsym{\beta}_r]_\ell}+\sumq \alphacq g_p(t-\taucq)e^{-\mathrm{j}2\pi \nucq pT}e^{-\mathrm{j}2\pi m[\bsym{\beta}_c]_q}\right) dt.
\end{align}
Substituting for $g_p(t)$ and computing the first integral produces
\begin{align}
&[\mathbf{\widehat{y}}_p(f)]_m\nonumber\\
=&\suml \alpharl  \widehat{s}(f)e^{-\mathrm{j}2\pi\left( \taurl f + \nurl pT + m[\bsym{\beta}_r]_\ell\right) } +\sumq \alphacq e^{-\mathrm{j}2\pi \left(\taucq f + \nucq pT + m[\bsym{\beta}_c]_q \right)} \sumk [\mathbf{g}_p]_k \int\limits_{pT}^{pT+T} e^{-\mathrm{j}2\pi (ft- k\Delta f t)}dt
\end{align}

Sampling uniformly at the rate 
$f_n = \frac{Bn}{M}=\Delta f n$, where $B$ is the signal bandwidth, $n=-N , \dots, N$, and $M = 2N+1$ is the number of frequency samples, yields

\begin{align}
    &[\mathbf{\widehat{y}}_p(f_n)]_m \nonumber\\
    &= \suml[\bsym{\alpha}_r]_\ell \widehat{s}(f_n)e^{-\mathrm{j}2\pi (n \Delta f\taurl+\pi\nurl pT+m[\bsym{\beta}_r]_\ell)} +\sumq [\bsym{\alpha}_c]_q e^{-\mathrm{j}2\pi n \Delta f\taucq }e^{-\mathrm{j}2\pi[\overline{\bsym{\nu}}_c]_{q}pT}\sumk [\mathbf{g}_p]_k\int\limits_{pT}^{pT+T} e^{\mathrm{j}2\pi \Delta f (k-n) t}dt,
    \label{ft_2}
\end{align}
where the number of frequency samples is equal to the number of subcarrier frequencies, i.e. $M=K$. 
Define entries of the vector $\mathbf{y}\in \mathbb{C}^{N_R MP\times 1}$ as  $\en{\mathbf{y}}_{\widetilde{m}} = [\widehat{\mathbf{y}}_p(f_n)]$, where the index $\widetilde{m}$ refers to a multidimensional linear indexing of the given by $\widetilde{m}=n+N+Mp+MPr$, with $\widetilde{m}=0,\dots, MPN_R, n=-N,\dots, N$, $p=0,\dots,P-1$ and $r=0,\dots, N_R-1$. Then, using the property $\int_{1}^{T}e^{-\mathrm{j}2\pi \Delta f(k-n)t}dt=0$ with $k \neq n$, all $n=-M,\dots,M$ samples. We have that
\begin{align}
    [\mathbf{y}]_{\widetilde{m}} = &\suml[\bsym{\alpha}_r]_\ell [\mathbf{s}]_n e^{-\mathrm{j}2\pi(n[{\bsym{\tau}}_r]_\ell+ p[{\bsym{\nu}}_r]_\ell +m[\bsym{\beta}_r]_\ell)} {+}\sumq [\bsym{\alpha}_c]_q[\mathbf{g}_p]_{n} e^{-\mathrm{j}2\pi(n[{\bsym{\tau}}_c]_q+ p[{\bsym{\nu}}_c]_q +m[\bsym{\beta}_c]_q)},
    \label{eq:y_p}
\end{align}
where $[{\bsym{\tau}}_r]_\ell = \frac{[\overline{\bsym{\tau}}_r]_\ell}{T} \in [0,1]$ and $[\bsym{\tau}_c]_\ell = \frac{[\overline{\bsym{\tau}}_c]_\ell}{T} \in [0,1]$ are the normalized delays; $[\bsym{\nu}_r]_\ell = \frac{[\overline{\bsym{\nu}}_r]_\ell}{\Delta f} \in [0,1]$ and 
$[\bsym{\nu}_c]_\ell = \frac{[\overline{\bsym{\nu}}_c]_\ell}{\Delta f} \in [0,1]$ are the normalized Doppler frequencies; the absolute values of  $[\bsym \alpha_r]_\ell$ and $[\bsym\alpha_c]_q$ are unit-norm after normalizing the signal by its magnitude, i.e. $\vert[\bsym \alpha_r]_\ell\vert = \vert[\bsym \alpha_c]_q\vert=1$; and $\en{\mathbf{s}}_{n+N} = S(f_n)$. 
\begin{remark}
    In general, OFDM communications employ a cyclic prefix (CP) for mitigating inter-symbol interference (ISI).  It is trivial to extend the M-DBD signal model to include CP; see. e.g., \cite[Remark 1]{vargas2023dual}.
\end{remark}
\begin{remark}
 The communications message $[\mathbf{g}_p]_k$ 
 is related to the constellation set (e.g., quadrature amplitude modulation) through a bijective map. In this paper, we adopt the symbol-agnostic approach of estimating only the communications waveform $[\mathbf{g}_p]_k$ rather than decoding a specific constellation, which would usually follow the waveform estimation step. These symbols are then recovered through \textit{conventional} blind decoding techniques. This aspect is, therefore, not a novel challenge because the key unknown DBD variable $[\mathbf{g}_p]_k$ is already been estimated before blind decoding. 
\end{remark}
\vspace{-12pt}
\subsection{Low-dimensional space representation}\label{sec:low_dimensional}
Our goal is to estimate the set of radar and communications parameters $\bsym{\alpha}_r$, $\bsym{\tau}_r$, $\bsym{\nu}_r$, $\bsym{\beta}_r$,  $\bsym{\alpha}_c$, $\bsym{\tau}_c$, $\bsym{\nu}_c$, and $\bsym{\beta}_c$, when the radar pulses $\mathbf{s}$ and communications symbols $\mathbf{g}$ are also unknown. In this inverse problem, the number of unknowns is $3LMPN_R(L+Q)$ and, therefore, the problem is highly ill-posed. 

Our strategy to solve this problem follows a method similar to the \textit{lifting trick} in BD problems \cite{chi2016guaranteed,yang2016super}. Herein,  $\mathbf{s}$ and $\mathbf{g}$ are assumed to lie in a given low-dimensional subspace, i.e., $\mathbf{s}=\mathbf{Tu}$, $\mathbf{g}=\mathbf{Dv}$, where $\mathbf{u}\in\mathbb{C}^{J}$ is the unknown coefficient vector of the radar waveform; $\mathbf{v} = [\mathbf{v}_1^T,\dots,\mathbf{v}_P^T]^T \in \mathbb{C}^{PJ}$ is a vector that contains the $P$ coefficient vectors of the communications messages such that $\mathbf{v}_p\in\mathbb{C}^{J}$; and the matrices $\mathbf{T} \in \mathbb{C}^{M\times J}$ and $\mathbf{D}\in \mathbb{C}^{MP\times PJ}$ are known random transformation matrices. Moreover, $\mathbf{D}=\operatorname{blockdiag}(\mathbf{D}_1,\dots,\mathbf{D}_P), \mathbf{D}_p\in\mathbb{C}^{M\times J}$ such that 
\begin{equation}
\mathbf{D} = 
\left[\begin{matrix}
\mathbf{D}_1& \mathbf{0} & \hdots & \mathbf{0}   \\
\mathbf{0} & \mathbf{D}_2 &  \hdots & \mathbf{0}   \\
\mathbf{0}&\mathbf{0} & \small{\ddots} & \vdots  \\
\mathbf{0} & \mathbf{0} & \hdots  & \mathbf{D}_P
\end{matrix}\right] \nonumber,
\end{equation}
The signal in \eqref{eq:y_p} now becomes
\begin{align}
    [\mathbf{y}]_{\widetilde{m}} &=\suml[\bsym{\alpha}_r]_\ell \mathbf{t}_n^H \mathbf{v} e^{-\mathrm{j}2\pi(n[{\bsym{\tau}}_r]_\ell+ p[{\bsym{\nu}}_r]_\ell +m[\bsym{\beta}_r]_\ell)} +\sumq [\bsym{\alpha}_c]_q \mathbf{d}_v^H\mathbf{v} e^{-\mathrm{j}2\pi(n[{\bsym{\tau}}_c]_q+ p[{\bsym{\nu}}_c]_q +m[\bsym{\beta}_c]_q)},
    \label{eq:y_p1}
\end{align}
where the index $v$ corresponds to the linear indexing $v = n + N +pM$, 
 $\mathbf{t}_n$ refers to the $n$-th column of $\mathbf{T}$ and $\mathbf{d}_v$ is the $v$-th column of $\mathbf{D}$. 
{Define vectors for the continuous-valued delay parameter $\mathbf{a(\tau)}_N = \big[e^{\mathrm{j}2\pi(\tau (-N))}, \dots, e^{\mathrm{j}2\pi(\tau (N)}\big]{^T} \in\mathbb{C}^{M}$, Doppler modulation $\mathbf{a(\nu)}_P = \big[e^{\mathrm{j}2\pi(\nu (1))}, \dots, e^{\mathrm{j}2\pi(\nu (P)}\big] {^T}\in\mathbb{C}^{P}$, and DoA $\mathbf{a(\beta)}_{N_R} = \big[e^{\mathrm{j}2\pi(\beta (1))}, \dots, e^{\mathrm{j}2\pi(\beta (N_R)}\big]{^T} \in\mathbb{C}^{N_R}$. Denote the vector $\mathbf{w}(\mathbf{r}) =\mathbf{a(\beta)}_{N_R}\otimes \mathbf{a(\nu)}_{P} \otimes \mathbf{a(\tau)}_{N} \in \mathbb{C}^{MPN_R}$ with $\mathbf{r} = [\bsym{\tau}_r,\bsym{\nu}_r,\bsym{\beta}_r]{^T}$. The channel vectors become $\mathbf{h}_r = \suml[\bsym{\alpha}_r]_\ell\mathbf{w(\mathbf{r}_\ell)}$ and $\mathbf{h}_c = \sumq[\bsym{\alpha}_c]_q\mathbf{w}(\mathbf{c}_q)$, where $\mathbf{c} = [\bsym{\tau}_c,\bsym{\nu}_c,\bsym{\beta}_c]{^T}$.} The observation vector becomes
\begin{equation}
    [\mathbf{y}]_{\widetilde{m}} = \mathbf{h}_r^H\mathbf{e}_{\widetilde{m}}\mathbf{t}_n\mathbf{u} + \mathbf{h}_c^H\mathbf{e}_{\widetilde{m}}\mathbf{d}_m\mathbf{v}, \label{eq:model_vect}
\end{equation}
where $\mathbf{e}_{\widetilde{m}}$ is the ${\widetilde{m}}$-th canonical vector of $\mathbb{R}^{MPN_R}$. Define the random sensing matrices $\mathbf{G}_{\widetilde{m}} = \mathbf{e}_{\widetilde{m}}\mathbf{t}_n^H \in \mathbb{C}^{MPN_R\times J}$ and  $\mathbf{A}_{\widetilde{m}} = \mathbf{e}_{\widetilde{m}}\mathbf{d}_m^H \in \mathbb{C}^{MPN_R\times PK}$. Denote $\mathbf{X}_r = \mathbf{u}\mathbf{h}_r^H \in \mathbb{C}^{J\times MPN_R}$ and $\mathbf{X}_c = \mathbf{v}\mathbf{h}_c^H$ as rank-one matrices that contain the unknown variables (channel parameters and signal coefficients). Then, the observation vector is a linear transformation of $\mathbf{X}_r$ and $\mathbf{X}_c$  
\begin{equation}
    \mathbf{y} = \mathcal{B}_r(\mathbf{X}_r)+\mathcal{B}_c(\mathbf{X}_c),
\end{equation}
where the linear operators $\mathcal{B}_r:\mathbb{C}^{J\times N_RMP}\rightarrow \mathbb{C}^{N_RMP}$ and $\mathcal{B}_c:\mathbb{C}^{PJ\times N_RMP}\rightarrow \mathbb{C}^{MPN_R}$ are defined as $[\mathcal{B}_r(\mathbf{X}_r)]_{\widetilde{m}}= \operatorname{Tr}(\mathbf{G}_{\widetilde{m}}\mathbf{X}_r)$ and $ [\mathcal{B}_c(\mathbf{X}_c)]_{\widetilde{m}} = \operatorname{Tr}(\mathbf{A}_{\widetilde{m}}\mathbf{X}_c)$.

Note that each matrix $\mathbf{X}_r$ and $\mathbf{X}_c$ is built upon a linear combination  $L$ and $Q$ atoms. These atoms are rank-one matrices $\mathbf{u}\mathbf{w}(\mathbf{r}_\ell)^T$ and $\mathbf{v}\mathbf{w}(\mathbf{c}_q)^T$ parametrized by the continuous-valued variables $\mathbf{r}_\ell$ and $\mathbf{c}_q$. Based on the atomic decomposition and the continuous-valued parametrization of the matrices $\mathbf{X}_r$ and $\mathbf{X}_c$, the ANM minimization is suitable to solve our DBD problem. 

Similar methods have been previously employed in super-resolution (SR) problems  \cite{barzegar2017estimation,chao2017semidefinite,wu2018high}. However, the transmit waveforms were known in these studies and only DoA was the unknown parameter to be estimated. A joint continuous-valued delay, Doppler, and DoA SR estimation was considered in \cite{heckel2016superMIMO} but the transmit signals were known \textit{a priori}. The blind 3-D SR is investigated in \cite{suliman2019exact} but with only one signal. Moreover, none of these studies 
considered the superposition of radar and communications signals. 

\vspace{-10pt}
\section{Recovery via SoMAN Minimization}
\label{sec:formulation}
The radar and communications channels are characterized by a few $L+Q\ll MPN_R$ continuous-valued parameters. To leverage upon the sparsity of channels, we adopt the SoMAN framework \cite{vargas2023dual} in 3-D. 
Define the sets of atoms for the radar and communications signals, respectively, as
\begin{align}
    &\mathcal{A}_r = \Big\{\mathbf{u}\mathbf{w}(\mathbf{r})^H: \mathbf{r}\in[0,1]^3,||\mathbf{u}||_2 = 1   \Big\},
\end{align}    
and
\begin{align}
    &\mathcal{A}_c = \Big\{\mathbf{v}\mathbf{w}(\mathbf{c})^H: \mathbf{c}\in[0,1]^3,||\mathbf{v}||_2 = 1   \Big\}.
    \label{eq:atomic_sets}
\end{align}
The corresponding atomic norms are
\begin{align}
    ||\mathbf{X}_r||_{\mathcal{A}_r} = \hspace{-4mm}\inf_{\stackrel{[\bsym{\alpha}_r]_\ell \in \mathbb{C}, \boldsymbol{r}_\ell \in [0,1]^3}{||\mathbf{u}||_2 = 1}} \Bigg\{\sum_\ell \vert[\bsym{\alpha}_r]_\ell\vert \Big| \mathbf{X}_r = \sum_\ell [\bsym{\alpha}_r]_\ell\mathbf{u}\mathbf{w}(\mathbf{r}_\ell)^H\Bigg\},
\end{align}    
and
\begin{align}
    ||\mathbf{X}_c||_{\mathcal{A}_c} = \hspace{-4mm}\inf_{\stackrel{[\bsym{\alpha}_c]_q\in \mathbb{C}, \boldsymbol{c}_q \in [0,1]^3}{||\mathbf{v}||_2 = 1}} \Bigg\{\sum_q \vert[\bsym{\alpha}_c]_q\vert\Big| \mathbf{X}_c = \sum_q [\bsym{\alpha}_c]_q\mathbf{v}\mathbf{w}(\mathbf{c}_q)^H\Bigg\}.
\end{align}
Then, our proposed 3-D SoMAN optimization is
\begin{align}
    &\minimize_{\mathbf{X}_r,\mathbf{X}_c} ||\mathbf{X}_r||_{\mathcal{A}_r} +||\mathbf{X}_c||_{\mathcal{A}_c} 
    \;\text{subject to }     \mathbf{y} = \mathcal{B}_r(\mathbf{X}_r) + \mathcal{B}_c(\mathbf{X}_c).
    \label{eq:primal_problem}
\end{align}

To formulate the SoMAN SDP, we resort to the following dual problem of \eqref{eq:primal_problem} obtained from the Lagrangian of the primal objective:
\begin{align}
    &\underset{\mathbf{q}}{\textrm{maximize}}\langle\mathbf{q,y}\rangle_{\mathbb{R}}
    \text{ subject to } \Vert\mathcal{B}_r^\star(\mathbf{q})\Vert^\star_{\mathcal{A}_r}\leq1, 
    \Vert \mathcal{B}_c^\star(\mathbf{q})\Vert^\star_{\mathcal{A}_c}\leq1, 
\label{eq:dual_problem_op}    
\end{align}
where $\mathcal{B}_r^\star: \mathbb{C}^{N_RMP}\rightarrow\mathbb{C}^{J \times N_RMP}$ and $\mathcal{B}_c^\star: \mathbb{C}^{N_RMP}\rightarrow\mathbb{C}^{PJ \times N_RMP}$ are adjoint operators of $\mathcal{B}_r$ and $\mathcal{B}_c$, respectively, such that $\mathcal{B}_r^\star(\boldsymbol{q}) = \sum_{m=1}^{N_R}\sump\sum_{n=-N}^{N}[\mathbf{q}]_{\widetilde{m}}\mathbf{G}_{\widetilde{m}}^H$ and $\mathcal{B}_c^\star(\mathbf{q}) = \sum_{m=1}^{N_R}\sump\sum_{n=-N}^{N}\mathbf{q}_{\widetilde{m}}\mathbf{A}_{\widetilde{m}}^H $. We observe that the constraints on the dual problem are, in fact, bounded vector-valued tri-variate trigonometric polynomials, or positive hyperoctant trigonometric polynomials (PhTPs), defined as
\begin{align}
    &\mathbf{f}_r(\mathbf{r}) =  \sum_{m=1}^{N_R}\sump\sum_{n=-N}^{N}[\mathbf{q}]_{\widetilde{m}}\mathbf{G}_{\widetilde{m}}^H\mathbf{w}(\mathbf{r}) \in \mathbb{C}^{J},
    \label{eq:poly_r}
\end{align}
and
\begin{align}
    &\mathbf{f}_c(\mathbf{c})=\sum_{m=1}^{N_R}\sump\sum_{n=-N}^{N}[\mathbf{q}]_{\widetilde{m}}\mathbf{A}_{\widetilde{m}}^H\mathbf{w}(\mathbf{c}) \in \mathbb{C}^{JP}.
    \label{eq:poly_c}
\end{align}

Following the bounded real lemma (BRL) and sum-of-squares relaxations \cite{dumitrescu2017positive}, PhTPs have a \textit{Gram} matrix parametrization that allows us to convert the inequalities in \ref{eq:dual_problem_op} to linear matrix inequalities (LMIs). This observation yields the following SDP of 3-D SoMAN minimization:
\begin{align}
    &\underset{\mathbf{q,Q}}{\textrm{maximize}}\quad \langle\mathbf{q,y}\rangle_{\mathbb{R}}\nonumber\\
    &\text{subject to }\mathbf{Q}\succeq 0,\nonumber\\&\hphantom{\text{subject to }}  
    \begin{bmatrix}
        \mathbf{Q} & \widehat{\mathbf{Q}}_r^H \\
        \widehat{\mathbf{Q}}_r & \mathbf{I}_J
        \end{bmatrix}
    \succeq0,\;
    \begin{bmatrix}
        \mathbf{Q} & \widehat{\mathbf{Q}}_c^H \\
        \widehat{\mathbf{Q}}_c & \mathbf{I}_{JP} 
        \end{bmatrix}\succeq 0,\;
    \text{Tr}(\boldsymbol{\Theta}_\mathbf{n}\mathbf{Q}) = \delta_{\mathbf{n}},\label{dual_opt}
\end{align}
where {$$\widehat{\mathbf{Q}}_r =  \sum_{m=1}^{N_R}\sump\sum_{n=-N}^{N}[\mathbf{q}]_{\widetilde{m}}\mathbf{G}_{\widetilde{m}}^H\in \mathbb{C}^{N_RMP\times J},$$ and $$ \widehat{\mathbf{Q}}_c =\sum_{m=1}^{N_R}\sump\sum_{n=-N}^{N}[\mathbf{q}]_{\widetilde{m}}\mathbf{A}_{\widetilde{m}}^H\in \mathbb{C}^{N_RMP\times PJ},$$} are the coefficients of 3-D PhTPs; the matrix 
$\boldsymbol{\Theta}_\mathbf{n} = \boldsymbol{\Theta}_{n_3} \otimes \boldsymbol{\Theta}_{n_2} \otimes \boldsymbol{\Theta}_{n_1}$;  $\bsym{\Theta}_n$ is Toeplitz with ones in the $n$-th diagonal with $0<n_1<m_1$, $-m_2<n_2<m_2$ and $-m_3<n_3<m_3$, such that $n_1,n_2$ and $n_3$ are degrees of the sum-of-squares relaxation, which defines the size of the \textit{Gram} matrix $\mathbf{Q}$; and 
$\delta_\mathbf{n} = 1 $ if $\mathbf{n} = [0,0,0]$ and $0$ otherwise. 
%
%
Later, we generalize this SDP to practical scenarios such as receiver antenna errors and the presence of noise (cf. Section \ref{sec:errors}. We mention a few other SDP extensions such as unsynchronized transmission, unequal number of pulses and messages, and n-tuple 3-D DBD in Section \ref{sec:summ}. 

\begin{remark}
To tackle the SDP, we use off-the-shelf solvers such as SDPT3 \cite{grant2009cvx}. However, this approach has cubic complexity order of the matrix $\mathbf{Q}$, here $\mathcal{O}(M^3P^3N_R^3)$, which becomes infeasible for large signal dimensions. In this case, one may employ low-complexity approaches that have been proposed for ANM-based recovery such as reweighting the continuous dictionary \cite{mingjiu2022gridless}, accelerating proximal gradient  \cite{wang2018ivdst}, or employing prior knowledge to apply block iterative $\ell_1$ minimization \cite{cho2015block}. In our recent work \cite{monsalve2023beurling}, we solved the 1-D DBD via a low-rank matrix Hankel recovery approach but this technique remains unexamined for multiple variables. 
\end{remark}
\vspace{-12pt}
\subsection{Recovery of transmit signals}
Computing the dual polynomials in \eqref{eq:poly_r} and \eqref{eq:poly_c} yields the estimated channel parameters $\{\widehat{\mathbf{r}}_\ell\}_{\ell=1}^{L}, \{\widehat{\mathbf{c}}_q\}_{q=1}^{Q}$. Thereafter, the radar and communications coefficient vectors $\mathbf{v} $ and $\mathbf{u}$ are estimated by solving an over-determined linear system of equations. Define matrices $\mathbf{W}_r \in \mathbb{C}^{MPN_R\times LJ}$ and $\mathbf{W}_c \in \mathbb{C}^{MPN_R\times PQJ}$ as 
\begin{equation*}
\mathbf{W}_r = 
\left[\begin{array}{ccc}
\mathbf{w}\left(\widehat{\mathbf{r}}_{1}\right)^{H} \mathbf{A}_{1} & \ldots & \mathbf{w}\left(\widehat{\mathbf{r}}_{L}\right)^{H}  \mathbf{A}_{1} \\
\vdots & \ddots & \vdots \\
\mathbf{w}\left(\widehat{\mathbf{r}}_{1}\right)^{H} \mathbf{A}_{MPN_R} & \ldots & \mathbf{w}\left(\widehat{\mathbf{r}}_{L}\right)^{H}  \mathbf{A}_{MPN_R}
\end{array}\right],
\end{equation*}
and
\begin{equation*}
\mathbf{W}_c = 
\left[\begin{array}{ccc}
\mathbf{w}\left(\widehat{\mathbf{c}}_{1}\right)^{H} \mathbf{G}_{1} & \ldots & \mathbf{w}\left(\widehat{\mathbf{c}}_{Q}\right)^{H}  \mathbf{G}_{1} \\
\vdots & \ddots & \vdots \\
\mathbf{w}\left(\widehat{\mathbf{c}}_{1}\right)^{H} \mathbf{G}_{MPN_R} & \ldots & \mathbf{w}\left(\widehat{\mathbf{c}}_{Q}\right)^{H}  \mathbf{G}_{MPN_R}
\end{array}\right].
\end{equation*}
Denote the vector containing the  desired coefficient vectors as \begin{equation}
    \mathbf{p} = \left[[\bsym{\alpha}_c]_1\mathbf{u},\dots,[\bsym{\alpha}_r]_L\mathbf{u}^T,[\bsym{\alpha}_c]_1\mathbf{v}^T,\dots, [\bsym{\alpha}_c]_Q\mathbf{v}^T\right]^T, \label{eq:p_vector}
\end{equation} and define the matrix 
\begin{equation}
    \mathbf{W} = [\mathbf{W}_r,\mathbf{W}_c] \label{eq:w_matrix}.
\end{equation}
The coefficient vector is then recovered (up to a scaling factor) by the least squares solution of the optimization problem
\begin{equation}
    \minimize_\mathbf{p}\Vert \mathbf{W}\mathbf{p}- \mathbf{y}\Vert_2
\end{equation}
Once the coefficient vectors $\widehat{\mathbf{u}}$ and $\widehat{\mathbf{v}}$ are estimated, the original transmit signals are computed using the basis matrices, i.e., $\widehat{\mathbf{s}} = \mathbf{T}\widehat{\mathbf{u}}$  and $\widehat{\mathbf{g}} = \mathbf{D}\widehat{\mathbf{v}}$.

\subsection{Performance analysis}
\label{sec:perf}
The following Proposition~\eqref{prop:dual_cert} states the conditions for the exact recovery of radar and communications channel parameters.  
\begin{proposition}
\label{prop:dual_cert}
 Denote the sets of radar and communications channel parameters by $\mathcal{R} = \{\mathbf{r}_\ell\}_{\ell=0}^{L-1}$ and $\mathcal{C} = \{\mathbf{c}_q\}_{q=0}^{Q-1}$, respectively. The solutions of \eqref{eq:primal_problem} are $\widehat{\mathbf{X}}_r$ and $\widehat{\mathbf{X}}_c$. Then, $\widehat{\mathbf{X}}_r={\mathbf{X}}_r$ and $\widehat{\mathbf{X}}_c={\mathbf{X}}_c$ are the optimal solutions of \eqref{eq:primal_problem} if there exist two 3-D PhTPs such that
\begin{align}
    \mathbf{f}_r(\mathbf{r}_\ell) &= \mathrm{sign}( \alpharl) \mathbf{u} \hspace{1em} \text{if} \hspace{1em} \forall \mathbf{r}_\ell \in \mathcal{R}, \label{eq:cert_1}\\
    \mathbf{f}_c(\mathbf{c}_q) &= \mathrm{sign}( \alphacq) \mathbf{v} \hspace{1em} \text{if} \hspace{1em} \forall\mathbf{c}_q \in \mathcal{C},  \label{eq:cert_2}\\
    \Vert \mathbf{f}_r(\mathbf{r}) \Vert_2^2 &< 1 \hspace{1em } \forall \mathbf{r} \in [0,1]^3 \setminus \mathcal{R},  \label{eq:cert_3}\\
    \Vert \mathbf{f}_c(\mathbf{c}) \Vert_2^2  &< 1 \hspace{1em }  \forall \mathbf{c} \in [0,1]^3 \setminus \mathcal{C},\label{eq:cert_4}
\end{align}
where $\operatorname{sign}(c) = \frac{c}{|c|}$.
\end{proposition}
\begin{proof}
If the variable $\mathbf{q}$ is dual feasible, we have 
\begin{flalign}
    &\langle\mathbf{q,y}\rangle_{\mathbb{R}} = \langle\mathcal{B}_r^*(\mathbf{q}),\mathbf{X}_r\rangle_{\mathbb{R}}+ \langle\mathcal{B}_c^*(\mathbf{q}),\mathbf{X}_c\rangle_{\mathbb{R}}\nonumber\\
    & = \suml\alpharl^* \langle\mathcal{B}_r^*(\mathbf{q}),\mathbf{u}\mathbf{w}(\mathbf{r}_\ell)^H\rangle_{\mathbb{R}} + \sumq \alphacq^* \langle\mathcal{B}_c^*(\mathbf{q}),\mathbf{v}\mathbf{w}(\mathbf{c}_q)^H\rangle_{\mathbb{R}}\nonumber \\
    &= \suml\alpharl^* \langle\mathbf{f}_r(\mathbf{r}_\ell),\mathbf{u}\rangle_{\mathbb{R}} + \sumq \alphacq^* \langle\mathbf{f}_c(\mathbf{c}_q),\mathbf{v}\rangle_{\mathbb{R}}\nonumber\\
    & =\suml\alpharl^*\textrm{sign}( \alpharl) + \sumq [\bsym{\alpha}_c]_q^*\textrm{sign}( [\bsym{\alpha}_c]_q)\nonumber\\
    &= \suml |[\bsym{\alpha}_r]_\ell| + \sumq |[\bsym{\alpha}_c]_q|\geq ||\mathbf{X}_r||_{\mathcal{A}_r} + ||\mathbf{X}_c||_{\mathcal{A}_c}.\label{eq:lower_bound_dual}
\end{flalign}
On the other hand, it follows from H\"{o}lder inequality that
\begin{align}
\langle\mathbf{q,y}\rangle_{\mathbb{R}}&=\langle\mathcal{B}_r^*(\mathbf{q}),\mathbf{X}_r\rangle_{\mathbb{R}}+ \langle\mathcal{B}_c^*(\mathbf{q}),\mathbf{X}_c\rangle_{\mathbb{R}}\nonumber\\
&\leq ||\mathcal{B}_r^*(\mathbf{q})||_{\mathcal{A}_r}^*||\mathbf{X}_r||_{\mathcal{A}_r} + ||\mathcal{B}_c^*(\mathbf{q})||_{\mathcal{A}_c}^*||\mathbf{X}_c||_{\mathcal{A}_c}\nonumber\\
&\leq ||\mathbf{X}_r||_{\mathcal{A}_r}+||\mathbf{X}_c||_{\mathcal{A}_c},
\label{eq:upper_bound_dual}
\end{align}
where the first inequality results from Cauchy-Schwarz inequality and the last inequality follows from \eqref{eq:cert_1}-
\eqref{eq:cert_4}. Therefore, based on \eqref{eq:lower_bound_dual} and \eqref{eq:upper_bound_dual}, we conclude that $\langle\mathbf{q,y}\rangle_{\mathbb{R}}=||\mathbf{X}_r||_{\mathcal{A}_r}+||\mathbf{X}_c||_{\mathcal{A}_c}$. Hence, the pair $(\mathbf{X}_r,\mathbf{X}_c)$ is primal optimal and, from strong duality, $\mathbf{q}$ is dual optimal.
\end{proof}

We now state our main theoretical result in the following Theorem~\ref{th:main}, which stipulates the minimum number of samples required for perfect recovery. Recall the following useful properties. 
\begin{definition}[Isotropy \cite{candes2011probabilistic}] The distribution $\mathcal{F}$ is isotropic if  given a random vector $\mathbf{x} \in \mathbb{C}^N \sim \mathcal{F}$ satisfies the following condition: 
    \begin{equation}
    \label{eq:isotropy}
        \mathbb{E}[\mathbf{x}\mathbf{x}^H] =\mathbf{I}_N.
    \end{equation}
   which means that the columns have unit variance and are uncorrelated.
\end{definition}
\begin{definition}[Incoherence \cite{candes2011probabilistic}] The distribution $\mathcal{F}$ is incoherent if given a vector $\mathbf{x}=[x_0,\dots,x_{N-1}] \sim \mathcal{F}$ and given coherence parameter $\mu$ it satisfies that:
\begin{equation}
\max_{0<i<N-1}|x_i|^2\leq\mu.
    \end{equation}
\end{definition}


\begin{theorem}
\label{th:main}
Assume the columns of $\mathbf{T}$ and $\mathbf{D}$  are drawn from a distribution that satisfies the isotropy and incoherence properties. Further, if $\Vert\mathbf{v}\Vert_2^2=\Vert\mathbf{u}\Vert_2^2=1$ and $\vert\bsym{\alpha}_r\vert = \vert\bsym{\alpha}_c\vert=1$, then there exists a numerical constant $C$ such that
\begin{align*}
    MPN_R &\geq  C\mu \operatorname{max}(L,Q)J\log^2\left(\frac{MPN_R J}{\delta}\right)\times\operatorname{max}\left\{\log\left(\frac{MPN_RQJ}{\delta}\right),\log\left(\frac{MPN_RLJ}{\delta}\right)\right\}, 
\end{align*}
are the number of samples required to guarantee that, with a probability of $1-\delta$, the pair of matrices $\{\mathbf{X}_r,\mathbf{X}_c\}$ is the minimizer of  \eqref{eq:primal_problem}.
\end{theorem}
\begin{proof}
    See Section~\ref{sec:proof}.
\end{proof}


\vspace{-10pt}
\section{Noise and Steering Vector Errors}
\label{sec:errors}
We consider some practical scenarios for SoMAN-based recovery of unknown parameters in 3-D DBD. In the presence of additive Gaussian noise $\bsym{\omega} \sim \mathcal{N}(0,\sigma_{\bsym{\omega}}) \in \mathbb{R}^{MPN_R}$, steering vector gain error $\mathbf{n} \in \mathbb{C}^{N_R}$, and phase error $\bsym{\phi}\in \mathbb{R}^{N_R}$, the received signal in \eqref{eq:model_vect} becomes
\begin{equation}
      [\mathbf{y}]_{\widetilde{m}} = (1+[\mathbf{n}]_m)e^{\mathrm{j}[\bsym{\phi}]_m}\left(\mathbf{h}_r^H\mathbf{e}_{\widetilde{m}}\mathbf{t}_n\mathbf{u} + \mathbf{h}_c^H\mathbf{e}_{\widetilde{m}}\mathbf{d}_m\mathbf{v}\right) + [\bsym{\omega}]_m.
\end{equation}

\subsection{SDP formulation}
 Define the diagonal matrix $\mathbf{E} = \operatorname{diag}(\mathbf{e}) \in \mathbb{C}^{N_R\times N_R}$ where the entries of $\mathbf{e}$ are $[\mathbf{e}]_{m} = (1+[{\mathbf{n}}]_{{m}})e^{\mathrm{j}\bsym{\phi}_{{m}}} -1 , m= 1,\dots, N_R$. Consider the matrix $\widetilde{\mathbf{E}} = \mathbf{E} \otimes \mathbf{I}_{MP}\in \mathbb{C}^{MPN_R\times MPN_R}$ that encapsulates the gain an dphase errors. Then, we employ the matrices {$$\widetilde{\mathbf{X}}_r = \mathbf{u}\mathbf{h}_r^H(\widetilde{\mathbf{E}} + \mathbf{I}_{MPN_R}),$$ and $$\widetilde{\mathbf{X}}_c = \mathbf{v}\mathbf{h}_c^H(\widetilde{\mathbf{E}} + \mathbf{I}_{MPN_R}),$$} in the corresponding linear sensing model $\mathbf{y} = \mathcal{B}_r(\widetilde{\mathbf{X}}_r)+\mathcal{B}_c(\widetilde{\mathbf{X}}_c)$. 
 
 Assume that the error vector is bounded $\Vert\mathbf{e}\Vert_2 \leq \varepsilon_e$ and, hence, the matrix norm $\Vert{\mathbf{E}}\Vert\leq \varepsilon_e$. Following the the property $\Vert\mathbf{A}\otimes\mathbf{B} \Vert = \Vert\mathbf{A}\Vert\Vert\mathbf{B}\Vert$, we have $\Vert\widetilde{\mathbf{E}}\Vert\leq\varepsilon_e$. 
 Consider the following, respectively, radar and communications atomic norms that are now weighted by the gain/phase errors:
\begin{align}
    ||\widetilde{\mathbf{X}}_r||_{\widetilde{\mathcal{A}}_r}= \inf_{\stackrel{[\bsym{\alpha}_r]_\ell \in \mathbb{C}, \boldsymbol{r}_\ell \in [0,1]^3}{||\mathbf{u}||_2 = 1,\Vert\widetilde{\mathbf{E}}\Vert\leq \varepsilon_e}} \Bigg\{\sum_\ell \vert[\bsym{\alpha}_r]_\ell\vert \Big| \widetilde{\mathbf{X}}_r = \sum_\ell [\bsym{\alpha}_r]_\ell\mathbf{u}\mathbf{w}(\mathbf{r}_\ell)^H (\widetilde{\mathbf{E}} + \mathbf{I}_{MPN_R})\Bigg\}\label{eq:an_weighted_rad}
    \end{align}
    and
    \begin{align}
   ||\widetilde{\mathbf{X}}_c||_{\widetilde{\mathcal{A}}_c} = \inf_{\stackrel{\bsym{\alpha}_c[q]\in \mathbb{C}, \boldsymbol{c}_q \in [0,1]^3}{||\mathbf{v}||_2 = 1,\Vert\widetilde{\mathbf{E}}\Vert\leq \varepsilon_e}} \Bigg\{\sum_q \vert[\bsym{\alpha}_c]_q\vert\Big|  \widetilde{\mathbf{X}}_c =& \sum_q [\bsym{\alpha}_c]_q\mathbf{v}\mathbf{w}(\mathbf{c}_q)^H(\widetilde{\mathbf{E}} + \mathbf{I}_{MPN_R})\Bigg\}\label{eq:an_weighted_com}.
\end{align}

To tackle noise, we add the regularization or penalty term to the primal problem in \eqref{eq:primal_problem}:
\begin{equation}
    \minimize_{\widetilde{\mathbf{X}}_r,\widetilde{\mathbf{X}}_c} \Vert \mathbf{y} - \mathcal{B}_r(\widetilde{\mathbf{X}}_r) - \mathcal{B}_c(\widetilde{\mathbf{X}}_c)  \Vert_2 + \mu_r \Vert\widetilde{\mathbf{X}}_r\Vert_{\widehat{\mathcal{A}}_r} + \mu_c \Vert\widetilde{\mathbf{X}}_c\Vert_{\widehat{\mathcal{A}}_c},\label{eq:primar_errors}
\end{equation}
where $\mu_r$ and $\mu_c$ are the regularization parameters. 
In the sequel, we also derive the optimal value of these parameters depending on the noise level $\sigma_{\bsym{\omega}}$ and gain/phase error vector bound. 

To formulate the dual problem of \eqref{eq:primar_errors}, consider the corresponding Lagrangian function:
\begin{align}
    \mathcal{L}(\widetilde{\mathbf{X}}_r,\widetilde{\mathbf{X}}_r,\mathbf{q},\mathbf{x}) = &\Vert\mathbf{x}-\mathbf{y}\Vert + (\mu_r\Vert\widetilde{\mathbf{X}}_r\Vert_{\widetilde{\mathcal{A}}_r}+\mu_c\Vert\mathbf{\widetilde{X}}_c\Vert_{\widetilde{\mathcal{A}}_c})+\langle\mathbf{x}-\mathcal{B}(\widetilde{\mathbf{X}}_r)_r-\mathcal{B}(\widetilde{\mathbf{X}}_c)_c,\mathbf{q}\rangle.\nonumber
\end{align}
The dual problem is 
\begin{equation}
\maximize_{\mathbf{q}}\min_{\widetilde{\mathbf{X}}_r,\widetilde{\mathbf{X}}_c,\mathbf{x}} \mathcal{L}(\widetilde{\mathbf{X}}_r,\widetilde{\mathbf{X}}_c,\mathbf{q},\mathbf{x}).\nonumber
\end{equation}
This is equivalent to
\begin{equation}
\maximize_{\mathbf{q}}\; \{\mathcal{L}_1(\mathbf{q})-\mathcal{L}_2(\mathbf{q})\},\nonumber
\end{equation}
where . i.e.,{
\begin{equation}
    \mathcal{L}_1(\mathbf{q}) = \min_\mathbf{x}\frac{1}{2}\Vert\mathbf{x-y}\Vert_2^2+\langle\mathbf{x},\mathbf{q}\rangle= \Vert\mathbf{q}-\mathbf{y}\Vert_2^2  + \frac{1}{2}\Vert\mathbf{y}\Vert_2^2,\nonumber
\end{equation}
and  
\begin{align}
    \mathcal{L}_2(\mathbf{q}) &=\max_{\widetilde{\mathbf{X}}_r,\widetilde{\mathbf{X}}_c}\langle\mathcal{B}(\widetilde{\mathbf{X}}_r)_r+\mathcal{B}(\widetilde{\mathbf{X}}_c)_c,\mathbf{q}\rangle -(\mu_r\Vert\widetilde{\mathbf{X}}_r\Vert_{\mathcal{A}_r}+\mu_c\Vert\widetilde{\mathbf{X}}_c\Vert_{\mathcal{A}_c})\nonumber\\&= \max_{\widetilde{\mathbf{X}}_r,\widetilde{\mathbf{X}}_c}(\langle\widetilde{\mathbf{X}}_r,\mathcal{B}^*_r(\mathbf{q})\rangle-\mu_r\Vert\widetilde{\mathbf{X}}_r\Vert_{\mathcal{A}_r})+(\langle\widetilde{\mathbf{X}}_c,\mathcal{B}^*_c(\mathbf{q})\rangle-\mu_c\Vert\widetilde{\mathbf{X}}_c\Vert_{\mathcal{A}_c}).\nonumber
\end{align}}
Further, following the definition of the dual norm,
\begin{equation}
    \mathcal{L}_2(\mathbf{q}) = \max_{\widetilde{\mathbf{X}}_r,\widetilde{\mathbf{X}}_c}(I_r(\Vert\mathbf{q}\Vert^*_{\widetilde{\mathcal{A}}_r}\leq \mu_r)+I_c(\Vert\mathbf{q}\Vert^*_{\mathcal{A}_c}\leq \mu_c)),\nonumber
\end{equation}
where 
\begin{equation}
    I_r(\Vert\mathbf{q}\Vert^*_{\widehat{\mathcal{A}}_r}\leq \mu_r) =  \left\{\begin{array}{cc}
        0 & \text{if }  \Vert\mathbf{q}\Vert^*_{\widetilde{\mathcal{A}}_r}\leq \mu_r, \\
         \infty & \text{otherwise,} \\
    \end{array}\right. \nonumber
\end{equation}
and
\begin{equation}
    I_c(\Vert\mathbf{q}\Vert^*_{\mathcal{A}_c}\leq \mu_c) =  \left\{\begin{array}{cc}
        0 & \text{if }  \Vert\mathbf{q}\Vert^*_{\mathcal{A}_c}\leq \mu_c, \\
         \infty & \text{otherwise,} \\
    \end{array}\right. \nonumber
\end{equation}
are indicator functions. The dual problem becomes
\begin{align}
    &\underset{\mathbf{q}}{\textrm{minimize}}\Vert\mathbf{q-y}\Vert_2
    \text{ subject to } \Vert\mathcal{B}_r^\star(\mathbf{q})\Vert^\star_{\widehat{\mathcal{A}}_r}\leq\mu_r, 
    \Vert \mathcal{B}_c^\star(\mathbf{q})\Vert^\star_{\widehat{\mathcal{A}}_c}\leq\mu_c, 
\label{eq:dual_problem_op_errors}    
\end{align}

The following Proposition~\ref{prop:sdp_error} states the SDP of \eqref{eq:dual_problem_op_errors}.
\begin{proposition}\label{prop:sdp_error}

    For $\Vert\widetilde{\mathbf{E}}\Vert\leq \varepsilon_e$, the SDP formulation of \eqref{eq:dual_problem_op_errors} is
    \begin{subequations}
 \begin{align}
    &\underset{\mathbf{q,Q}}{\textrm{minimize}}\quad \Vert\mathbf{q-y}\Vert_2\nonumber\\
    &\text{subject to }\mathbf{Q}\succeq 0,\;
    \begin{bmatrix}
        \mathbf{Q} & \widehat{\mathbf{Q}}_r^H \\
        \widehat{\mathbf{Q}}_r & \mu_r\mathbf{I}_J
        \end{bmatrix} 
    \succeq0,\;
    \begin{bmatrix}
        \mathbf{Q} & \widehat{\mathbf{Q}}_c^H \\
        \widehat{\mathbf{Q}}_c & \mu_c\mathbf{I}_{JP} 
        \end{bmatrix}\succeq 0,\;
    \label{eq:lmis_error}\\&\hphantom{\text{subject to }}
    \operatorname{Tr}(\boldsymbol{\Theta}_\mathbf{n}\mathbf{Q}) =0 \quad(\mathbf{n}\neq\mathbf{0})\label{eq:const3_error}\\&\hphantom{\text{subject to }}
    \operatorname{Tr}(\mathbf{Q}) + (\varepsilon_e+2\sqrt{MPN_R})\varepsilon_e\Vert \mathbf{Q}\Vert_2 -1\leq0\label{eq:dual_opt_sdp_error}
\end{align}
\end{subequations}
\end{proposition} 
\begin{proof}
Define the dual-modified atomic norms of \eqref{eq:an_weighted_rad} and \eqref{eq:an_weighted_com} as, respectively,
\begin{align}
\left\|\mathcal{B}_r^{*}(\mathbf{q})\right\|_{\mathcal{A}_r}^*&=\operatorname{sup}\left|\langle\mathbf{u}\mathbf{w}(\mathbf{r})^H(\widetilde{\mathbf{E}}+\mathbf{I}_{MPN_R}), \mathcal{B}_r^*(\mathbf{q})\rangle\right\vert\nonumber\\
&=\operatorname{sup}\left|\mathbf{u}^H\ \mathcal{B}_r^*(\mathbf{q})^H\mathbf{w}_r(\mathbf{r})^H(\mathbf{E}+\mathbf{I}_{MPN_R})\right|\nonumber\\
&=\operatorname{sup}\left\Vert \mathcal{B}_r^*(\mathbf{q})^H\left(\mathbf{w}_r(\mathbf{r})^H+\widetilde{\mathbf{e}}_r\right)\right\Vert, \label{eq:dual_norm_error_radar}
\end{align}
and
\begin{equation}
    \left\|\mathcal{B}_c^{*}(\mathbf{q})\right\|_{\mathcal{A}_c}^* = \sup \left\Vert\mathcal{B}_c^{*}(\mathbf{q}) (\mathbf{w}_c(\mathbf{c})^H + \widetilde{\mathbf{e}}_c)\right\Vert, \label{eq:dual_norm_error_comms}
\end{equation}
where $\widetilde{\mathbf{e}}_r=\mathbf{w}_r(\mathbf{r})\widetilde{\mathbf{E}}$ and $\widetilde{\mathbf{e}}_c=\mathbf{w}_c(\mathbf{c})\widetilde{\mathbf{E}}$. 
The expression of the dual atomic norm contains PhTPs. Hence, we change the constraints $ \left\|\mathcal{B}_r^{*}(\mathbf{q})\right\|_{\mathcal{A}_r}^*\leq \mu_r$ and $ \left\|\mathcal{B}_c^{*}(\mathbf{q})\right\|_{\mathcal{A}_c}^*\leq \mu_c$ to LMIs. For instance, the radar LMI is
\begin{equation}
     \begin{bmatrix}
        \mathbf{Q} & \widehat{\mathbf{Q}}_r^H \\
        \widehat{\mathbf{Q}}_r & \mu_r\mathbf{I}_J
        \end{bmatrix}
    \succeq0,
\end{equation}
where the matrix $\mathbf{Q}$ is semidefinite positive. Using the Schur complement, we have $\mathbf{Q}- \widehat{\mathbf{Q}}_r^H  \frac{1}{\mu_r^2}\mathbf{I}_J\widehat{\mathbf{Q}}_r \succeq 0$. The communications LMI is similarly defined.

 For any vector $\boldsymbol{\upsilon}\in\mathbb{C}^{MPN_R}$ we have  $\boldsymbol{\upsilon}^H\mathbf{Q}\boldsymbol{\upsilon}\geq \frac{1}{\mu_r^2}\boldsymbol{\upsilon}\widehat{\mathbf{Q}}_r\mathbf{I}_{J}\widetilde{\mathbf{Q}}_r^H\boldsymbol{\upsilon}^H$. Thus, $\Vert\widehat{\mathbf{Q}}_r\boldsymbol{\upsilon}\Vert_2\leq \mu_r^2\boldsymbol{\upsilon}^H\mathbf{Q}\boldsymbol{\upsilon}$. 
 For $\boldsymbol{\upsilon} = \mathbf{w}_r(\mathbf{r}) + \widetilde{\mathbf{e}}_r$, then
 \begin{align}
     &\Vert \widetilde{\mathbf{Q}}_r ( \mathbf{w}_r(\mathbf{r}) + \widetilde{\mathbf{e}}_r) \Vert_2^2 \leq  \mu_r^2( \mathbf{w}_r(\mathbf{r}) + \widetilde{\mathbf{e}}_r) ^H\mathbf{Q} ( \mathbf{w}_r(\mathbf{r}) + \widetilde{\mathbf{e}}_r)\nonumber\\ &=\mu_r^2(\widetilde{\mathbf{e}}_r^H\mathbf{Q}\widetilde{\mathbf{e}}_r+2\langle \widetilde{\mathbf{e}}_r,\mathbf{Q}\mathbf{w}_r(\mathbf{r})\rangle_{\mathbb{R}} + \mathbf{w}_r(\mathbf{r})^H\mathbf{Q}\mathbf{w}_r(\mathbf{r}))\nonumber\\&\leq \mu_r^2(\varepsilon_e^2\Vert \mathbf{Q}\Vert_2+2\varepsilon_e\Vert\mathbf{Q}\mathbf{w}_r(\mathbf{r})\Vert_2 + \mathbf{w}_r(m)^H)\nonumber\\&\leq\mu_r^2(\varepsilon_e\Vert\mathbf{Q}\Vert_2(\varepsilon_e + 2\sqrt{MPN_R})+\mathbf{w}_r(\mathbf{r})^H\mathbf{Q}\mathbf{w}_r(\mathbf{r})),
 \end{align}
 where the first inequality follows from the quadratic identity $\mathbf{a}^H\mathbf{B}\mathbf{a} = \langle \mathbf{a},\mathbf{B}\mathbf{a}\rangle \leq \Vert\mathbf{B}\Vert_2 \Vert\mathbf{a}\Vert_2^2$\cite{martos1969subdefinite}. 
 The second inequality follows from $\Vert\mathbf{Q}\mathbf{w}_r(\mathbf{r})\Vert_2 \leq \sqrt{MPN_R}\Vert\mathbf{Q}\Vert_2$. Then, $\mathbf{Q}$ satisfies \eqref{eq:const3_error} and \eqref{eq:dual_opt_sdp_error}.

\end{proof}
\vspace{-12pt}
\subsection{Performance Analyses}
To find the optimal values of the regularization parameters in \eqref{eq:lmis_error}, 
we employ the following criteria based on the expected value of the dual norm to select the hyperparameters: 
\begin{equation}
    \mu_r = \rho\expec{\Vert\mathcal{B}_r^*(\bsym{\omega})\Vert_{\widetilde{A}_r}^*}, \mu_c = \rho\expec{\Vert\mathcal{B}_c(\bsym{\omega})\Vert_{\widetilde{A}_c}^*},
\end{equation}
with $\rho\geq 1$. Then, the following Theorem~\ref{thm:upperbound} establishes an upper bound for $\expec{\Vert\mathcal{B}_r(\bsym{\omega})\Vert_{\widetilde{A}_r}^*}$ and $\expec{\Vert\mathcal{B}_c(\bsym{\omega})\Vert_{\widetilde{A}_c}^*}$. \begin{theorem}
\label{thm:upperbound}
    When the transformation bases $\mathbf{T}$ and $\mathbf{D}$ achieve the incoherence and isotropy properties, the upper bounds on the dual weighted atomic norms \eqref{eq:dual_norm_error_radar} and \eqref{eq:dual_norm_error_comms} in the presence of additive Gaussian noise $\bsym{\omega}\sim \mathcal{N}(0,\sigma_{\bsym{\omega}}^2)$, satisfy
    \begin{align}    \expec{\Vert\mathcal{B}_r^*(\bsym{\omega})\Vert_{\widetilde{A}_r}^*} \leq \sigma_{\bsym{\omega}}\left(\varepsilon_e MPN_R + \Vert\mathbf{T}\Vert_F\sqrt{MPJN_R\log(MPN_R)}\right), \label{eq:mu_r}
      \end{align}
and
        \begin{align}   \expec{\Vert\mathcal{B}_c^*(\bsym{\omega})\Vert_{\widetilde{A}_c}^*} \leq \sigma_{\bsym{\omega}}\left(\varepsilon_eMPN_R + \Vert\mathbf{D}\Vert_F\sqrt{MP^2JN_R\log(MPN_R)}\right). \label{eq:mu_c}
    \end{align}
\end{theorem}
\begin{proof}
Following the definition of the dual atomic norm with antenna errors in \eqref{eq:dual_norm_error_radar}, we have
\begin{align*}
    \expec{\Vert\mathcal{B}_r^*(\bsym{\omega})\Vert_{\widetilde{A}_r}^*} &= \expec{ \left\Vert\mathcal{B}_r^*(\bsym{\omega})^H\left(\mathbf{w}_r(\mathbf{r})^H+\widetilde{\mathbf{e}}\right)\right\Vert}\\&\leq \expec{\left\Vert \mathcal{B}_r^*(\bsym{\omega})^H\mathbf{w}_r(\mathbf{r})^H\right\Vert} + \expec{\left\Vert \mathcal{B}_r^*(\bsym{\omega})^H\widetilde{\mathbf{e}}\right\Vert}.
\end{align*}
We bound the first term following \cite[Lemma V.I]{li2019atomic}, which employs matrix Bernstein inequality (MBI) to exploit the zero-mean of the noise and the incoherence of the representation basis, leading to
\begin{equation}
    \expec{\left\Vert \mathcal{B}_r^*(\bsym{\omega})^H\mathbf{w}_r(\mathbf{r})^H\right\Vert} \leq C\sigma_{\bsym{\omega}}\Vert\mathbf{T}\Vert_F\sqrt{MPN_RJ\log(MPN_R)}.\label{eq:bound_1}
\end{equation}
Consider now the definition of the adjoint operator $\mathcal{B}_r^*(\bsym{\omega}) = \sum_{m=1}^{N_R}\sum_{p=1}^{P} \sum_{n=-N}^{N} [\bsym\omega]_{\widetilde{m}}\mathbf{A}_{\widetilde{m}}$. Since $\Vert\widetilde{\mathbf{e}}\Vert\leq\varepsilon_e$, we have
\begin{align}
    \expec{\left\Vert \mathcal{B}_r^*(\bsym{\omega})^H\widetilde{\mathbf{e}}\right\Vert} &\leq \varepsilon_e\expec{\left\Vert \sum_{m=1}^{N_R}\sum_{p=1}^{P} \sum_{n=-N}^{N} [\bsym\omega]_{\widetilde{m}}\mathbf{A}_{\widetilde{m}} \right\Vert} \nonumber\\& \leq \expec{ \sum_{m=1}^{N_R}\sum_{p=1}^{P} \sum_{n=-N}^{N} [\bsym\omega]_{\widetilde{m}}\left\Vert\mathbf{e}_{\widetilde{m}}^H\mathbf{t}_n^H\mathbf{t}_n\mathbf{e}_{\widetilde{m}} \right\Vert}\nonumber\\&\leq \varepsilon_eMPN_R\sigma_{\bsym\omega}. \label{eq:bound_2}
\end{align}
Combining \eqref{eq:bound_1} and \eqref{eq:bound_2} completes the proof for $\mu_r$. The proof follows, \textit{mutatis mutandis}, for $\mu_c$.
\end{proof}

Our results are closely related to the theory of atomic norm denoising \cite{li2019atomic,bhaskar2013atomic,suliman2021mathematical}. However, these prior studies did not consider multiple antennas and, therefore, excluded the effect of steering vector gain/phase errors. In 3-D DBD, these errors result in the dual atomic norm containing two terms that each account for the channel parameters and errors. While \cite{chen2020new} included steering vector errors, they considered the DoA-only SR problem. As a result, their dual atomic norm depends on only the noise distribution. In our case, it depends on the adjoint operators $\mathcal{B}_r^*$ and $\mathcal{B}_r^*$. 
The following Proposition~\ref{prop:dual_cert_errors} provides the dual certificate of the SDP stated in Proposition~\ref{prop:sdp_error}.
\begin{proposition}
\label{prop:dual_cert_errors}
 Denote the sets $\mathcal{R} = \{\mathbf{r}_\ell\}_{\ell=0}^{L-1}$ and $\mathcal{C} = \{\mathbf{c}_q\}_{q=0}^{Q-1}$. The solutions of \eqref{eq:primar_errors} are $\widehat{\mathbf{X}}_r$ and $\widehat{\mathbf{X}}_c$. 
 if there exist two 3-D PhTPs $\mathbf{f}_r(\mathbf{r})$ and $\mathbf{f}_r(\mathbf{r})$ such that 
\begin{align}
    \Vert\mathbf{f}_r(\mathbf{r}_\ell)\Vert_2^2 &= a_r\hspace{1em} \text{if} \hspace{1em} \forall \mathbf{r}_\ell \in \mathcal{R}, \label{eq:cert_1_mod}\\
    \Vert\mathbf{f}_c(\mathbf{c}_q)\Vert_2^2 &= a_c\hspace{1em} \text{if} \hspace{1em} \forall\mathbf{c}_q \in \mathcal{C},  \label{eq:cert_2_mod}\\
    \Vert \mathbf{f}_r(\mathbf{r}) \Vert_2^2 &< a_r\hspace{1em} \text{if} \hspace{1em} \forall \mathbf{r}_\ell \in \mathcal{R} \hspace{1em } \forall \mathbf{r} \in [0,1]^3 \setminus \mathcal{R},  \label{eq:cert_3_mod}\\
    \Vert \mathbf{f}_c(\mathbf{c}) \Vert_2^2  &< a_c\hspace{1em} \text{if} \hspace{1em} \forall \mathbf{r}_\ell \in \mathcal{R} \hspace{1em }  \forall \mathbf{c} \in [0,1]^3 \setminus \mathcal{C},\label{eq:cert_4_mod}
\end{align}
where $a_r=\mu_r^2(1-(\varepsilon_e + 2\sqrt{MPN_R}) \epsilon_e\Vert \mathbf{Q}\Vert_2)$ and $a_c = \mu_c^2(1-(\varepsilon_e + 2\sqrt{MPN_R}) \epsilon_e\Vert \mathbf{Q}\Vert_2)$.
\end{proposition}
\begin{proof}
    Note that $\mathbf{f}_r(\mathbf{r}) = \widehat{\mathbf{Q}}_r\mathbf{w}_r(\mathbf{r})$  and $\mathbf{f}_c(\mathbf{c}) = \widehat{\mathbf{Q}}_c\mathbf{w}_c(\mathbf{c})$. Using the radar LMI in \eqref{eq:dual_problem_op_errors} and Schur complement, we get \begin{align*}
        &\mathbf{Q}\succeq0,\\
        &\mathbf{Q} - \mu_r^2\widehat{\mathbf{Q}}\mathbf{I}_{J}\widetilde{\mathbf{Q}}^H\succeq 0.
    \end{align*}
 For any vector $\mathbf{k}\in\mathbb{C}^{MPN_R}$ we have  $\mathbf{k}^H\mathbf{Q}\mathbf{k}\geq \mu_r^2\mathbf{k}\widehat{\mathbf{Q}}_r\mathbf{I}_{J}\widetilde{\mathbf{Q}}_r^H\mathbf{k}^H$. Thus, $\Vert\widehat{\mathbf{Q}}_r\mathbf{k}\Vert_2\leq \mu_r^2\mathbf{k}^H\mathbf{Q}\mathbf{k}$.  When $\mathbf{k} = \mathbf{w}_r(\mathbf{r})$, then using 
 \begin{align}
    &\operatorname{Tr}(\boldsymbol{\Theta}_\mathbf{n}\mathbf{Q}) =0 \quad(\mathbf{n}\neq\mathbf{0}), \nonumber\\&
    \operatorname{Tr}(\mathbf{Q}) + (\varepsilon_e+2\sqrt{MPN_R})\varepsilon_e\Vert \mathbf{Q}\Vert_2 -1\leq0,\nonumber
 \end{align}
 produces
 \begin{equation*}
 \Vert\widehat{\mathbf{Q}} \mathbf{w}_r(\mathbf{r})\Vert_2^2 \leq \mu_r^2(1- (\varepsilon_e+2\sqrt{MPN_R})\varepsilon_e\Vert \mathbf{Q}\Vert_2).
 \end{equation*}
 The proof follows, \textit{mutatis mutandis}, for the communications polynomial.\end{proof}
\subsection{Waveform recovery}
\label{subsec:wf_rec_error}
Define $\widehat{\mathbf{r}}$ and $\widehat{\mathbf{c}}$ to be the estimated radar and communications channel parameters obtained through the SDP in \eqref{eq:dual_opt_sdp_error}. To recover the coefficient vectors of radar and communications transmit signals, solve the optimization problem 
\begin{align}
    &\minimize_{\mathbf{v},\mathbf{u},\mathbf{e}}   \sum_{m=0}^{N_R}\sum_{p=1}^{P}\sum_{n=-N}^{N}\left\Vert[\mathbf{y}]_{\widetilde{m}} - [\mathbf{e}]_m\left(\sum_{\ell =1}^{L}\alpharl \mathbf{w}(\widehat{\mathbf{r}})^H\mathbf{e}_{\widetilde{m}}\mathbf{t}_n^H\mathbf{u} \right.\right.\left.\left.+\sum_{q =1}^{Q}\alphacq \mathbf{w}(\widehat{\mathbf{c}})^H\mathbf{e}_{\widetilde{m}}\mathbf{d}_v^H\mathbf{v}\right)\right\Vert_2\nonumber\\ 
    &\text{subject to } \Vert \mathbf{e}\Vert_2 = \varepsilon_e.&\label{eq:opt_erros}
\end{align} 
We solve this problem by alternating optimizing the variables $\mathbf{e}$, $\mathbf{u}$, and $\mathbf{v}$. Then, the Lagrangian of \eqref{eq:opt_erros} is
\begin{align}
    \mathcal{L}(\mathbf{v},\mathbf{u},\mathbf{e}) =    \sum_{m=1}^{N_R}&\left\Vert\mathbf{y}_{m} - [\mathbf{e}]_m \mathbf{z}_{m}\right\Vert_2^2 + \lambda (\Vert \mathbf{e}\Vert_2 - \varepsilon_e),\label{eq:lagrangian_errors}
\end{align}
where $\mathbf{z}_m = [[\mathbf{z}_m]_1,\dots,[\mathbf{z}_m]_{MP}] \in \mathbb{C}^{MP}$ and 
{ $$[\mathbf{z}_m]_v = \sum_{\ell =1}^{L}\alpharl \mathbf{w}(\widehat{\mathbf{r}})^H\mathbf{e}_{\widetilde{m}}\mathbf{t}_n^H\mathbf{u}+\sum_{q =1}^{Q}\alphacq \mathbf{w}(\widehat{\mathbf{c}})^H\mathbf{e}_{\widetilde{m}}\mathbf{d}_v^H\mathbf{v},$$} recall that $v=n+N+Mp$, and $v=1,\dots,MP$. The find $\mathbf{e}$ we solve 
 \begin{equation}
     \widehat{\mathbf{e}} = \argmin_{\mathbf{e}} \mathcal{L} (\mathbf{v},\mathbf{u},\mathbf{e}).
 \end{equation}
We solve this by taking the derivative $\frac{\partial \mathcal{L} (\mathbf{v},\mathbf{u},\mathbf{e})}{\partial \mathbf{e}} = 0$ to yield
\begin{equation}
    \widehat{e}_m = \frac{\mathbf{z}_m^H\mathbf{y}_m}{\mathbf{z}_m^H\mathbf{z}_m  -\lambda},\label{eq:opt_e_errors}
\end{equation}

Then, to estimate the coefficient vectors $\mathbf{v}$ and $\mathbf{u}$, define the matrix $\widetilde{\mathbf{W}} = \mathbf{W}\widehat{\mathbf{E}}$, where the matrix $\widehat{\mathbf{E}} = \operatorname{diag}(\widehat{\mathbf{e}}) \otimes \mathbf{I}_{MP} \mathbb{C}^{MPN_R}$ and $\mathbf{W}$ is defined as in \eqref{eq:w_matrix}. Recall the definition of the vector $\mathbf{p}$ in \eqref{eq:p_vector}. The coefficient vectors are then obtained by the least squares solution of
\begin{equation}
    \minimize_{\mathbf{p}} \Vert \mathbf{y} - \mathbf{\widehat{W}}\mathbf{p} \Vert_2.\label{eq:opt_p_errors}
\end{equation}
We then iteratively compute the updates in equation \eqref{eq:opt_e_errors} and \eqref{eq:opt_p_errors} to recover $\mathbf{v}$ and $\mathbf{u}$.


\vspace{-10pt}
\section{Proof of Theorem \ref{th:main}}
\label{sec:proof}

The existence of dual  polynomials guarantees that the optimal solution of the primal problem \eqref{eq:primal_problem} is the pair $\{\mathbf{X}_r,\mathbf{X}_c\}$ based on the derivation of the dual certificate. We first find the polynomials $\mathbf{f}_r(\mathbf{r})$ and $\mathbf{f}_c(\mathbf{c})$ that satisfy the following conditions to ensure maximum modulus of the dual polynomial occurs at the true parameter values:
\begin{align}
&\mathbf{f}_r(\mathbf{r}_\ell) = \mathrm{sign}( [\bsym{\alpha}_r]_\ell) \mathbf{u}, &\forall \mathbf{r}_\ell \in \mathcal{R}, \label{eq:cert_opt_1}\\
& -\mathbf{f}_r^{(1,0,0)}(\mathbf{r}_\ell) = \mathbf{0}_{J\times 1}, &\forall \mathbf{r}_\ell \in \mathcal{R},\label{eq:cert_opt_2}\\
& -\mathbf{f}_r^{(0,1,0)}(\mathbf{r}_\ell) = \mathbf{0}_{J\times 1}, &\forall \mathbf{r}_\ell \in \mathcal{R},\label{eq:cert_opt_3}\\
& -\mathbf{f}_r^{(0,0,1)}(\mathbf{r}_\ell) = \mathbf{0}_{J\times 1}, &\forall \mathbf{r}_\ell \in \mathcal{R},\label{eq:cert_opt_4}\\
& \hspace{1em}\mathbf{f}_c(\mathbf{c}_q) = \mathrm{sign}( [\bsym{\alpha}_c]_q) \mathbf{v},  & \forall \mathbf{c}_q \in \mathcal{C},\label{eq:cert_opt_5}\\
& -\mathbf{f}_c^{(1,0,0)}(\mathbf{c}_q) = \mathbf{0}_{PJ\times 1}, &\forall \mathbf{c}_q \in \mathcal{C},\label{eq:cert_opt_6}\\
& -\mathbf{f}_c^{(0,1,0)}(\mathbf{c}_q) = \mathbf{0}_{PJ\times 1}, &\forall \mathbf{c}_q \in \mathcal{C}, \label{eq:cert_opt_7}\\
& -\mathbf{f}_c^{(0,0,1)}(\mathbf{c}_q) = \mathbf{0}_{PJ\times 1}, &\forall \mathbf{c}_q \in \mathcal{C}, \label{eq:cert_opt_8}
\end{align}
where $\mathbf{f}^{(m',n',k')}(\mathbf{r}) = \frac{\partial ^{m'}}{\partial \tau^{m'}}\frac{\partial ^{n'}}{\partial \nu^{m'}} \frac{\partial ^{k'}}{\partial \beta^{k'}}\mathbf{f}(\mathbf{r})$.  

Prior works \cite{heckel2016super,candes2014towards,chi2016guaranteed,tang2013compressed} build such polynomials for SR and BD problems through random kernels. For 3-D DBD, we introduce \textit{multiple} random kernels associated with the radar and communications signals. However, contrary to previous works, the non-trivial challenge in our DBD problem is that the dual variable $\mathbf{q}$ remains the same for the radar and communications polynomials. To this end, we first obtain an initial expression for $\mathbf{q}$ by solving the following weighted constrained minimization problem
\begin{align}
\minimize_{{\mathbf{q}_0}} & \hspace{1em}\| \mathbf{W} {\mathbf{q}_0} \|_{2}^{2} \nonumber\\
\text { subject to: } & \hspace{1em} \mathbf{f}_r(\mathbf{r}_\ell) = \mathrm{sign}( [\bsym{\alpha}_r]_\ell) \mathbf{u}, &\forall \mathbf{r}_\ell \in \mathcal{R}, \\
& -\mathbf{f}_r^{(1,0,0)}(\mathbf{r}_\ell) = \mathbf{0}_{J\times 1}, &\forall \mathbf{r}_\ell \in \mathcal{R},\\
& -\mathbf{f}_r^{(0,1,0)}(\mathbf{r}_\ell) = \mathbf{0}_{J\times 1}, &\forall \mathbf{r}_\ell \in \mathcal{R},\\
& -\mathbf{f}_r^{(0,0,1)}(\mathbf{r}_\ell) = \mathbf{0}_{J\times 1}, &\forall \mathbf{r}_\ell \in \mathcal{R},\\
& \hspace{1em}\mathbf{f}_c(\mathbf{c}_q) = \mathrm{sign}( [\bsym{\alpha}_c]_q) \mathbf{v},  & \forall \mathbf{c}_q \in \mathcal{C},\\
& -\mathbf{f}_c^{(1,0,0)}(\mathbf{c}_q) = \mathbf{0}_{PJ\times 1}, &\forall \mathbf{c}_q \in \mathcal{C},\\
& -\mathbf{f}_c^{(0,1,0)}(\mathbf{c}_q) = \mathbf{0}_{PJ\times 1}, &\forall \mathbf{c}_q \in \mathcal{C}, \\
& -\mathbf{f}_c^{(0,0,1)}(\mathbf{c}_q) = \mathbf{0}_{PJ\times 1}, &\forall \mathbf{c}_q \in \mathcal{C}, 
\label{eq:opt_wq0}
\end{align}
where {$\mathbf{W}=\text{diag}([\omega_1,\cdots,\omega_{MPN_R}])$} is a diagonal weighting matrix. Note that {the} first and fourth constraints are the same as \eqref{eq:cert_1} and \eqref{eq:cert_2}, while the remaining constraints are conditions necessary to satisfy \eqref{eq:cert_3} {and} \eqref{eq:cert_4}. Building upon the definition of the PhTPs in \eqref{eq:poly_r} and \eqref{eq:poly_c}, define the matrix $\mathbf{F}\in\mathbb{C}^{(4LJ+4QPJ)\times MPN_R}$: 
\begin{align}
\mathbf{F}=\left[\begin{matrix}
\mathbf{t}_{-N}\mathbf{e}_{1}^{H} \mathbf{w}_r\left(\mathbf{r}_{1}\right) & \cdots & \mathbf{t}_{N}\mathbf{e}_{MPN_R}^{H} \mathbf{w}_r\left(\mathbf{r}_{1}\right) \\
\vdots & \ddots & \vdots \\
\mathbf{t}_{-N}\mathbf{e}_{1}^{H} \mathbf{w}_r\left(\mathbf{r}_{L}\right)  & \hdots & \mathbf{t}_{N}\mathbf{e}_{MPN_R}^{H} \mathbf{w}_r\left(\mathbf{r}_{L}\right)  \\
-\mathrm{j}2\pi(-N)\mathbf{t}_{-N}\mathbf{e}_{1}^{H} \mathbf{w}_r\left(\mathbf{r}_{1}\right)& \hdots &-\mathrm{j}2\pi(N)\mathbf{t}_{N}\mathbf{e}_{MPN_R}^{H} \mathbf{w}_r\left(\mathbf{r}_{1}\right)\\
\vdots & \ddots & \vdots \\
-\mathrm{j}2\pi(-N)\mathbf{t}_{-N}\mathbf{e}_{1}^{H} \mathbf{w}_r\left(\mathbf{r}_{L}\right)& \hdots &-\mathrm{j}2\pi(N)\mathbf{t}_{N}\mathbf{e}_{MPN_R}^{H} \mathbf{w}_r\left(\mathbf{r}_{L}\right)\\
-\mathrm{j}2\pi(1)\mathbf{t}_{-N}\mathbf{e}_{1}^{H} \mathbf{w}_r\left(\mathbf{r}_{1}\right)& \hdots &-\mathrm{j}2\pi(P)\mathbf{t}_{N}\mathbf{e}_{MPN_R}^{H} \mathbf{w}_r\left(\mathbf{r}_{1}\right)\\
\vdots & \ddots & \vdots \\
-\mathrm{j}2\pi(1)\mathbf{t}_{-N}\mathbf{e}_{1}^{H} \mathbf{w}_r\left(\mathbf{r}_{L}\right)& \hdots &-\mathrm{j}2\pi(P)\mathbf{t}_{N}\mathbf{e}_{MPN_R}^{H} \mathbf{w}_r\left(\mathbf{r}_{L}\right)\\
-\mathrm{j}2\pi(1)\mathbf{t}_{-N}\mathbf{e}_{1}^{H} \mathbf{w}_r\left(\mathbf{r}_{1}\right)& \hdots &-\mathrm{j}2\pi(N_R)\mathbf{t}_{N}\mathbf{e}_{MPN_R}^{H} \mathbf{w}_r\left(\mathbf{r}_{1}\right)\\
\vdots & \ddots & \vdots \\
-\mathrm{j}2\pi(1)\mathbf{t}_{-N}\mathbf{e}_{1}^{H} \mathbf{w}_r\left(\mathbf{r}_{L}\right)& \hdots &-\mathrm{j}2\pi(N_R)\mathbf{t}_{N}\mathbf{e}_{MPN_R}^{H} \mathbf{w}_r\left(\mathbf{r}_{L}\right)\\
\mathbf{d}_{-NP}\mathbf{e}_{1}^{H} \mathbf{w}_c\left(\mathbf{r}_{1}\right) & \cdots & \mathbf{d}_{NP}\mathbf{e}_{MPN_R}^{H} \mathbf{w}_c\left(\mathbf{r}_{1}\right) \\
\vdots & \ddots & \vdots \\
\mathbf{d}_{-NP}\mathbf{e}_{1}^{H} \mathbf{w}_c\left(\mathbf{r}_{Q}\right) & \cdots & \mathbf{d}_{NP}\mathbf{e}_{MPN_R}^{H} \mathbf{w}_c\left(\mathbf{r}_{Q}\right)  \\
-\mathrm{j}2\pi(-N)\mathbf{d}_{-NP}\mathbf{e}_{1}^{H} \mathbf{w}_c\left(\mathbf{r}_{1}\right)& \hdots &-\mathrm{j}2\pi(N)\mathbf{d}_{NP}\mathbf{e}_{MPN_R}^{H} \mathbf{w}_c\left(\mathbf{r}_{1}\right)\\
\vdots & \ddots & \vdots \\
-\mathrm{j}2\pi(-N)\mathbf{d}_{-NP}\mathbf{e}_{1}^{H} \mathbf{w}_c\left(\mathbf{r}_{Q}\right)& \hdots &-\mathrm{j}2\pi(N)\mathbf{d}_{NP}\mathbf{e}_{MPN_R}^{H} \mathbf{w}_c\left(\mathbf{r}_{Q}\right)\\
-\mathrm{j}2\pi(1)\mathbf{d}_{-NP}\mathbf{e}_{1}^{H} \mathbf{w}_c\left(\mathbf{r}_{1}\right)& \hdots &-\mathrm{j}2\pi(P)\mathbf{d}_{NP}\mathbf{e}_{MPN_R}^{H} \mathbf{w}_c\left(\mathbf{r}_{1}\right)\\
\vdots & \ddots & \vdots \\
-\mathrm{j}2\pi(1)\mathbf{d}_{-NP}\mathbf{e}_{1}^{H} \mathbf{w}_c\left(\mathbf{r}_{Q}\right)& \hdots &-\mathrm{j}2\pi(P)\mathbf{d}_{NP}\mathbf{e}_{MPN_R}^{H} \mathbf{w}_c\left(\mathbf{r}_{Q}\right)\\
-\mathrm{j}2\pi(1)\mathbf{d}_{-NP}\mathbf{e}_{1}^{H} \mathbf{w}_c\left(\mathbf{r}_{1}\right)& \hdots &-\mathrm{j}2\pi(N_R)\mathbf{d}_{NP}\mathbf{e}_{MPN_R}^{H} \mathbf{w}_c\left(\mathbf{r}_{1}\right)\\
\vdots & \ddots & \vdots \\
-\mathrm{j}2\pi(1)\mathbf{d}_{-NP}\mathbf{e}_{1}^{H} \mathbf{w}_c\left(\mathbf{r}_{Q}\right)& \hdots &-\mathrm{j}2\pi(N_R)\mathbf{d}_{NP}\mathbf{e}_{MPN_R}^{H} \mathbf{w}_c\left(\mathbf{r}_{Q}\right)
\end{matrix}\right].\nonumber
\end{align}
where the vectors $\mathbf{t}_n$, $\mathbf{w}_r(\mathbf{r})$, $\mathbf{w}_c(\mathbf{c})$, and $\mathbf{d}_v$ are as defined in Section \ref{sec:low_dimensional}.Denote the vector {$\mathbf{p}= [\mathbf{p}_r^T, \mathbf{p}_c^T]^T \in \mathbb{C}^{(4LK + 4QPK)\times 1}$}. where
\begin{align}
    \mathbf{p}_r &= [ \textrm{sign}(\bsym{\alpha}_r[1])\mathbf{u}^T,\cdots, \textrm{sign}(\bsym{\alpha}_r[L])\mathbf{u}^T,\mathbf{0}_{J\times 1}^T,\cdots,\mathbf{0}_{J\times 1}^T]^T, \nonumber \\
    \mathbf{p}_c &= [ \textrm{sign}(\bsym{\alpha}_c[1])\mathbf{v}^T,\cdots, \textrm{sign}(\bsym{\alpha}_c[Q])\mathbf{v}^T,\mathbf{0}_{PJ\times 1}^T,\cdots,\mathbf{0}_{PJ\times 1}^T]^T. \nonumber\\   
    \label{eq:RHS_dual_certificate}
\end{align}
Then, the optimization problem in \eqref{eq:opt_wq0} becomes
\begin{align}
\minimize_{{\mathbf{q}_0}} & \hspace{1em}\| \mathbf{W}{\mathbf{q}_0} \|_{2}^{2} \nonumber\\
\text{subject to } & \hspace{1em}\mathbf{F} {\mathbf{q}_0}=\mathbf{p}.
\label{eq:dual_first_v2}
\end{align}
Following the Karush–Kuhn–Tucker (KKT) conditions of the optimization problem \eqref{eq:dual_first_v2}, its least-squares solution is ${\mathbf{q}_0}= \left(\mathbf{W}^H\mathbf{W}\right)^{-1}\mathbf{F}^H \bsym{\lambda}$ where 
\begin{align*}
    &\bsym{\lambda} = [\bsym{\beta}^T,\bsym{\gamma}^T,\bsym{\zeta}^T,\bsym{\varpi}^T,\bsym{\eta}^T,\bsym{\theta}^T,\bsym{\xi}^T,\bsym{\varrho}^T]^T\\
    &\bsym{\beta} = [\bsym{\beta}_1^T,\cdots,\bsym{\beta}_{L}^T]^T,
    \bsym{\gamma} = [\bsym{\gamma}_1^T,\cdots,\bsym{\gamma}_{L}^T]^T,\\
    &\bsym{\zeta} = [\bsym{\zeta}_1^T,\cdots,\bsym{\zeta}_{L}^T]^T,\bsym{\varpi} = [\bsym{\varpi}_1,\dots,\bsym{\varpi}_{L}],  \bsym{\beta}_\ell,\bsym{\gamma}_\ell,\bsym{\zeta}_\ell ,\bsym{\varpi}_\ell\in \mathbb{C}^{J\times1}\\
    &\bsym{\eta} = [\bsym{\eta}_1^T,\cdots,\bsym{\eta}_{Q}^T]^T,
    \bsym{\theta} = [\bsym{\theta}_1^T,\cdots,\bsym{\theta}_{Q}^T]^T,\\
    &\bsym{\xi} = [\bsym{\xi}_1^T,\cdots,\bsym{\xi}_{Q}^T]^T, \bsym{\varrho} = [\bsym{\varrho}_1^T,\cdots,\bsym{\varrho}_{Q}^T]^T, \bsym{\eta}_q,\bsym{\theta}_q,\bsym{\xi}_q,\bsym{\varrho}_q \in \mathbb{C}^{PJ\times1} .  
\end{align*}
such that the product $\boldsymbol{\varsigma}=\mathbf{F}^H \bsym{\lambda}$ becomes
\begin{align}
\boldsymbol{\varsigma}= &\suml\left[\begin{smallmatrix}
\mathbf{w} \left(\mathbf{r}_{\ell}\right)^{H} \mathbf{e}_{1}\mathbf{t}_{-N}^{H}  \\
\vdots \\
\mathbf{w}_r\left(\mathbf{r}_{\ell}\right)^{H}\mathbf{e}_{MPN_R}\mathbf{t}_{N}^H
\end{smallmatrix}\right]\bsym{\beta}_\ell +\left[\begin{smallmatrix}
\mathrm{j}2\pi(-N)\mathbf{w}_r\left(\mathbf{r}_{\ell}\right)\mathbf{t}_{-N}\mathbf{e}_{1}^{H}  \\
\vdots \\
\mathrm{j}2\pi(N)\mathbf{w}_r\left(\mathbf{r}_{\ell}\right)\mathbf{t}_{N}\mathbf{e}_{MPN_R}^{H} 
\end{smallmatrix}\right]\bsym{\gamma}_\ell \nonumber\\&
+ \left[\begin{smallmatrix}
\mathrm{j}2\pi(1)\mathbf{w}_r\left(\mathbf{r}_{\ell}\right)\mathbf{t}_{-N}\mathbf{e}_{1}^{H}   \\
\vdots \\
\mathrm{j}2\pi(P)\mathbf{w}_r\left(\mathbf{r}_{\ell}\right)\mathbf{t}_{N}\mathbf{e}_{MPN_R}^{H} 
\end{smallmatrix}\right]\bsym{\zeta}_\ell+ \left[\begin{smallmatrix}
\mathrm{j}2\pi(1)\mathbf{w}_r\left(\mathbf{r}_{\ell}\right)\mathbf{t}_{-N}\mathbf{e}_{1}^{H}   \\
\vdots \\
\mathrm{j}2\pi(N_R)\mathbf{w}_r\left(\mathbf{r}_{\ell}\right)\mathbf{t}_{N}\mathbf{e}_{MPN_R}^{H} 
\end{smallmatrix}\right]\bsym{\varpi}_\ell+\nonumber\\ &
\sumq\left[\begin{smallmatrix}
\mathbf{w} \left(\mathbf{c}_{q}\right)^{H} \mathbf{e}_{1}\mathbf{d}_{-NP}^{H}  \\
\vdots \\
\mathbf{w}_c\left(\mathbf{r}_{q}\right)^{H}\mathbf{e}_{MPN_R}\mathbf{d}_{NP}^H
\end{smallmatrix}\right]\bsym{\eta}_q + 
\left[\begin{smallmatrix}
\mathrm{j}2\pi(-N)\mathbf{w}_c\left(\mathbf{r}_{q}\right)\mathbf{d}_{-NP}\mathbf{e}_{1}^{H}   \\
\vdots \\
\mathrm{j}2\pi(N)\mathbf{w}_c\left(\mathbf{r}_{q}\right)\mathbf{d}_{NP}\mathbf{e}_{MPN_R}^{H} 
\end{smallmatrix}\right]\bsym{\theta}_q \nonumber\\&
+ \left[\begin{smallmatrix}
\mathrm{j}2\pi(1)\mathbf{w}_c\left(\mathbf{r}_{q}\right)\mathbf{d}_{-NP}\mathbf{e}_{1}^{H}   \\
\vdots \\
\mathrm{j}2\pi(P)\mathbf{w}_c\left(\mathbf{r}_{q}\right)\mathbf{d}_{NP}\mathbf{e}_{MPN_R}^{H} 
\end{smallmatrix}\right]\bsym{\xi}_q \left[\begin{smallmatrix}
\mathrm{j}2\pi(1)\mathbf{w}_c\left(\mathbf{r}_{q}\right)\mathbf{d}_{-NP}\mathbf{e}_{1}^{H}   \\
\vdots \\
\mathrm{j}2\pi(N_R)\mathbf{w}_c\left(\mathbf{r}_{q}\right)\mathbf{d}_{NP}\mathbf{e}_{MPN_R}^{H} 
\end{smallmatrix}\right]\bsym{\varrho}_q. 
\label{eq:first_dual_v3}
\end{align}

Replacing {the expression of $\mathbf{q}_0$ in \eqref{eq:poly_r} leads to}
\begin{align}
    &\mathbf{f}_r(\mathbf{r})=\sum_{\ell=0}^{L-1}\left[\left(\sum_{\widetilde{m}} \frac{1}{\omega_{\widetilde{m}}^2}\mathbf{t}_{n}\mathbf{e}_{\widetilde{m}}^{H} \mathbf{w}_r(\mathbf{r})^H \mathbf{w}_r(\mathbf{r}_\ell) \mathbf{e}_{\widetilde{m}}\mathbf{t}_{n}^H\right) \bsym{\beta}_{\ell} \right.\nonumber\\
    &\left.+\left(\sum_{\widetilde{m}} \mathrm{j}2\pi(n)\mathbf{t}_{n}\mathbf{e}_{\widetilde{m}}^{H} \mathbf{w}_r(\mathbf{r})^H \mathbf{w}_r\left(\mathbf{r}_{\ell}\right)\mathbf{e}_{\widetilde{m}}\mathbf{t}_{n}^H\right)\bsym{\gamma}_\ell  \right.\nonumber\\
    &\left.+\left(\sum_{\widetilde{m}} \mathrm{j}2\pi(p)\mathbf{t}_{n}\mathbf{e}_{\widetilde{m}}^{H} \mathbf{w}_r(\mathbf{r})^H \mathbf{w}_r\left(\mathbf{r}_{\ell}\right)\mathbf{e}_{\widetilde{m}}\mathbf{t}_{n}^H\right) \bsym{\zeta}_\ell\right.\nonumber\\
    &\left.+\left(\sum_{\widetilde{m}} \mathrm{j}2\pi(m)\mathbf{t}_{n}\mathbf{e}_{\widetilde{m}}^{H} \mathbf{w}_r(\mathbf{r})^H \mathbf{w}_r\left(\mathbf{r}_{\ell}\right)\mathbf{e}_{\widetilde{m}}\mathbf{t}_{n}^H\right) \bsym{\varpi}_\ell\right]\nonumber\\
    &+\sum_{q=0}^{Q-1}\frac{1}{\omega_{\widetilde{m}}^2}\left[\left(\sum_{\widetilde{m}} \mathbf{t}_{n}\mathbf{e}_{\widetilde{m}}^{H} \mathbf{w}_r(\mathbf{c})^H \mathbf{w}(\mathbf{c}_q) \mathbf{e}_{\widetilde{m}}\mathbf{d}_{v}^H\right) \bsym{\eta}_{q}\right.\nonumber\\
    &\left.+\left(\sum_{\widetilde{m}} \mathrm{j}2\pi(n)\mathbf{t}_{n}\mathbf{e}_{\widetilde{m}}^{H} \mathbf{w}_r(\mathbf{r})^H \mathbf{w}_r\left(\mathbf{r}_{\ell}\right)\mathbf{e}_{\widetilde{m}}\mathbf{d}_{v}^H\right)\bsym{\theta}_\ell  \right.\nonumber\\
    &\left.+\left(\sum_{\widetilde{m}} \mathrm{j}2\pi(p)\mathbf{t}_{n}\mathbf{e}_{\widetilde{m}}^{H} \mathbf{w}_r(\mathbf{r})^H \mathbf{w}_r\left(\mathbf{r}_{\ell}\right)\mathbf{e}_{\widetilde{m}}\mathbf{d}_{v}^H\right) \bsym{\xi}_q\right.\nonumber\\
     &\left.+\left(\sum_{\widetilde{m}} \mathrm{j}2\pi(m)\mathbf{t}_{n}\mathbf{e}_{\widetilde{m}}^{H} \mathbf{w}_r(\mathbf{r})^H \mathbf{w}_r\left(\mathbf{r}_{\ell}\right)\mathbf{e}_{\widetilde{m}}\mathbf{d}_{v}^H\right) \bsym{\varrho}_q\right],\label{eq:fr_r_1}
\end{align}
where $\mathbf{r} = [\tau,\nu,\beta]$ and $\mathbf{r}_\ell = [\tau_\ell,\nu_\ell,\beta_\ell]$ are as mentioned in \eqref{eq:rad_channel1} and $\mathbf{w}_r(\cdot)$ as in Section~\ref{sec:low_dimensional}. Substituting
\begin{align}
  \mathbf{b}_n\mathbf{e}_{\widetilde{m}}^H\mathbf{w}(\mathbf{r})&=e^{\mathrm{j}2\pi(n\tau + p \nu+r\beta)}\mathbf{b}_n, \\
 \mathbf{w}(\mathbf{r})^H   \mathbf{e}_{j}\mathbf{b}_n^H&=e^{-\mathrm{j}2\pi(n\tau + p \nu+r\beta)}\mathbf{b}_n^H,
\end{align}
in \eqref{eq:fr_r_1} produces 
\begin{align}
    &\mathbf{f}_r(\mathbf{r})=\sum_{\ell=1}^{L}\sum_{\widetilde{m}}\frac{1}{\omega_{\widetilde{m}}^2}\left[     e^{\mathrm{j}2\pi n(\tau -\tau_\ell)}e^{\mathrm{j}2\pi p(\nu - \nu_\ell)}e^{\mathrm{j}2\pi\widetilde{m}(\beta - \beta_\ell)}\mathbf{t}_{n}\mathbf{t}_{n}^H \bsym{\beta}_{\ell} \right.\left.+ \mathrm{j}2\pi(n)e^{\mathrm{j}2\pi n(\tau-\tau_\ell)}e^{\mathrm{j}2\pi p(\nu - \nu_\ell)}e^{\mathrm{j}2\pi\widetilde{m}(\beta - \beta_\ell)}\mathbf{t}_{n}\mathbf{t}_{n}^H\bsym{\gamma}_\ell  \right.\nonumber\\
    &\left.+\mathrm{j}2\pi(p) e^{\mathrm{j}2\pi n(\tau-\tau_\ell)}e^{\mathrm{j}2\pi p(\nu - \nu_\ell)}e^{\mathrm{j}2\pi\widetilde{m}(\beta - \beta_\ell)}\mathbf{t}_{n}\mathbf{t}_{n}^H \bsym{\zeta}_\ell\right.\left.+\mathrm{j}2\pi(m) e^{\mathrm{j}2\pi n(\tau-\tau_\ell)}e^{\mathrm{j}2\pi p(\nu - \nu_\ell)}e^{\mathrm{j}2\pi\widetilde{m}(\beta - \beta_\ell)}\mathbf{t}_{n}\mathbf{t}_{n}^H \bsym{\varpi}_\ell\right]\nonumber\\
    &+\sum_{q=1}^{Q}\frac{1}{\omega_{\widetilde{m}}^2}\left[e^{\mathrm{j}2\pi n(\tau-\tau_q)}e^{\mathrm{j}2\pi\widetilde{m}(\nu - \nu_q)}e^{\mathrm{j}2\pi p(\beta - \beta_q)} \mathbf{t}_{n}\mathbf{d}_{v}^H \bsym{\eta}_{q}\right.\left.+ \mathrm{j}2\pi(n)e^{\mathrm{j}2\pi n(\tau-\tau_q)}e^{\mathrm{j}2\pi p(\nu - \nu_q)}e^{\mathrm{j}2\pi\widetilde{m}(\beta - \beta_q)}\mathbf{t}_{n}\mathbf{d}_{v}^H\bsym{\theta}_\ell  \right.\nonumber\\
    &\left.+ \mathrm{j}2\pi(p)e^{\mathrm{j}2\pi n(\tau-\tau_q)}e^{\mathrm{j}2\pi p(\nu - \nu_q)}e^{\mathrm{j}2\pi\widetilde{m}(\beta - \beta_q)}\mathbf{t}_{n}\mathbf{d}_{v}^H \bsym{\xi}_q\right.\left.+ \mathrm{j}2\pi(m)e^{\mathrm{j}2\pi n(\tau-\tau_q)}e^{\mathrm{j}2\pi p(\nu - \nu_q)}e^{\mathrm{j}2\pi\widetilde{m}(\beta - \beta_q)}\mathbf{t}_{n}\mathbf{d}_{v}^H \bsym{\varrho}_q\right]. \label{eq:fr_r_2} 
\end{align}
Define the random matrices 
\begin{align}
    \mathbf{M}\left(\mathbf{r}\right)=&\sum_{\widetilde{m}} \frac{1}{\omega_{\widetilde{m}}^2}e^{\mathrm{j}2\pi n\tau}e^{\mathrm{j}2\pi p\nu} e^{\mathrm{j}2\pi\widetilde{m}\beta}    \mathbf{t}_{n}\mathbf{t}_{n}^H,\;\in \mathbb{C}^{J\times K},
\end{align}
\begin{align}
    \mathbf{N}\left(\mathbf{r}\right)=&\sum_{\widetilde{m}} \frac{1}{\omega_{\widetilde{m}}^2}e^{\mathrm{j}2\pi n\tau}e^{\mathrm{j}2\pi p\nu}  e^{\mathrm{j}2\pi\widetilde{m}\beta}  \mathbf{t}_{n}\mathbf{d}_{v}^H,\;\in \mathbb{C}^{J\times PK}.
\end{align}
Then, 
\eqref{eq:fr_r_2} becomes
\begin{align}
   \mathbf{f}_r (\mathbf{r}) &= \sum_{\ell=1}^{L} \mathbf{M}(\mathbf{r-r}_\ell)\bsym{\beta}_\ell + \mathbf{M}^{(1,0,0)}(\mathbf{r-r}_\ell)\bsym{\gamma}_\ell  + \mathbf{M}^{(0,1,0)}(\mathbf{r-r}_\ell)\bsym{\zeta}_\ell + \mathbf{M}^{(0,0,1)}(\mathbf{r-r}_\ell)\bsym{\varpi}_\ell \nonumber\\&+\sum_{q=1}^{Q} {\mathbf{N}}(\mathbf{r- c}_q)\bsym{\eta}_q + {\mathbf{N}}^{(1,0,0)}(\mathbf{r-c}_q)\bsym{\theta}_q + {\mathbf{N}}^{(0,1,0)}(\mathbf{r-c}_q)\bsym{\xi}_q+{\mathbf{N}}^{(0,0,1)}(\mathbf{r-c}_q)\bsym{\varrho}_q,
    \label{eq:polyr_random_v1}
\end{align}
Choosing the weights $\omega_{\widetilde{m}} = \sqrt{\frac{N}{g_N(n)}} \sqrt{\frac{P}{g_P(p)}}\sqrt{\frac{N_R}{g_{N_R}(m)}}$, {and  the following vector $g_N=g_P=g_{N_R}$}
\begin{align}
    g_N(n) = \frac{1}{N}\sum_{k=\text{max}\{ n-N,-N\}}^{\text{min}\{ n+N,N\}} \left(1-\frac{\vert k\vert}{M}\right)\left(1-\frac{\vert n-k\vert}{M}\right),\nonumber
\end{align}
ensures that the matrices $\mathbf{M}$ and $\mathbf{N}$ are randomized $3$-D version of the squared Fej\'er kernel
\begin{align}
    K(\mathbf{r}) = K_N(\tau)K_P(\nu)K_{N_R}(\beta),
    \label{eq:kernel_2D}
\end{align}
where
\begin{align}
    K_N(\tau) &= \left(\frac{\text{sin}(T\pi \tau)}{T\text{sin}(\pi \tau)}\right)^4, \hspace{1em} T=\frac{N}{2}+1,\nonumber\\
    K_N(\tau) &= \sum_{n=-N}^{N} g_N(n) \textrm{e}^{\textrm{j}2\pi \tau n}.
\end{align}

 The communications dual polynomial, \textit{mutatis mutandis}, is 
\begin{align}
\mathbf{f}_c(\mathbf{c}) & = \sum_{\ell=1}^{L} \mathbf{N}(\mathbf{c-r}_\ell)^H\bsym{\beta}_\ell + \mathbf{N}^{(1,0,0)}(\mathbf{c-r}_\ell)^H\bsym{\gamma}_\ell + \mathbf{N}^{(0,1,0)}(\mathbf{c-r}_\ell)^H\bsym{\zeta}_\ell + \mathbf{N}^{(0,0,1)}(\mathbf{c-r}_\ell)^H\bsym{\varpi}_\ell\nonumber\\&+\sum_{q=1}^{Q} {\mathbf{P}}(\mathbf{c-c}_q)\bsym{\eta}_q + {\mathbf{P}}^{(1,0,0)}(\mathbf{c-c}_q)\bsym{\theta}_q + 
    {\mathbf{P}}^{(0,1,0)}(\mathbf{c-c}_q)\bsym{\xi}_q + {\mathbf{P}}^{(0,0,1)}(\mathbf{c-c}_q)\bsym{\varrho}_q,
    \label{eq:polyc_random_final}
\end{align}\normalsize
where 
\begin{align}
    \mathbf{P}\left(\mathbf{r}\right)=&\frac{1}{MPN_R}\sum_{\widetilde{m}} g_N(n) g_P(p) g_{N_R}(m) e^{\mathrm{j}2\pi n\tau}e^{\mathrm{j}2\pi p\nu}e^{\mathrm{j}2\pi\widetilde{m}\beta}    \mathbf{d}_{v}\mathbf{d}_{v}^H.
\end{align}

Next, define the vectors
\begin{align}
    \mathbf{j}_r = [\bsym{\beta}^T,{\kappa}\bsym{\gamma}^T,\kappa \bsym{\zeta}^T,\kappa\bsym{\varpi}^T]^T,\hspace{1em} \mathbf{j}_c = [\bsym{\eta}^T,{\kappa}\bsym{\theta}^T,\kappa \bsym{\xi}^T,\kappa\bsym{\varrho}^T]^T,
\end{align}
and matrices $\mathbf{H}_1 \in \mathbb{C}^{4LJ \times 4LJ}$, $\mathbf{H}_2 \in \mathbb{C}^{4LJ \times 4PQJ}$, and $\mathbf{H}_3 \in \mathbb{C}^{4PQJ \times 4PQJ}$ such that
\begin{align}
\mathbf{H}_i=\left[\begin{matrix}
{\mathbf{E}_i^{(0,0,0)}}& \frac{1}{\kappa} {\mathbf{E}}_i^{(1,0,0)} & \frac{1}{\kappa} {\mathbf{E}}_i^{(0,1,0)}&{\mathbf{E}}_i^{(0,0,1)}\\
-\frac{1}{\kappa} {\mathbf{E}}_i^{(1,0,0)} & -\frac{1}{\kappa^{2}} {\mathbf{E}}_i^{(2,0,0)} & -\frac{1}{\kappa^{2}} {\mathbf{E}}_i^{(1,1,0)}& {\mathbf{E}}_i^{(1,0,1)}\\
-\frac{1}{\kappa} {\mathbf{E}}_i^{(0,1,0)} & -\frac{1}{\kappa^{2}} {\mathbf{E}}_i^{(1,1,0)} & -\frac{1}{\kappa^{2}} {\mathbf{E}}_i^{(0,2,0)}&-\frac{1}{\kappa^{2}} {\mathbf{E}}_i^{(0,1,1)}\\-\frac{1}{\kappa} {\mathbf{E}}_i^{(0,0,1)} & -\frac{1}{\kappa^{2}} {\mathbf{E}}_i^{(1,0,1)} & -\frac{1}{\kappa^{2}} {\mathbf{E}}_i^{(0,1,1)}&-\frac{1}{\kappa^{2}} {\mathbf{E}}_i^{(0,0,2)}
\end{matrix}\right],\;i = 1,2,3,
\end{align}
where $\kappa = \sqrt{\varphi''(0)}$. The matrix ${\mathbf{E}_1}^{(m',n',k')} \in \mathbb{C}^{LJ \times LJ}$ comprises a total of 
$L^2$ block matrices, each of size $J\times J$, as
\begin{align}
{\mathbf{E}_1}^{\left(m^{\prime}, n^{\prime},k^{\prime}\right)}=\begin{bmatrix}
\mathbf{M}^{\left(m^{\prime}, n^{\prime},k^{\prime}\right)} \left(\mathbf{r}_{1}-\mathbf{r}_{1}\right) & \ldots & \mathbf{M}^{\left(m^{\prime}, n^{\prime},k^{\prime}\right)}\left(\mathbf{r}_{1}- \mathbf{r}_{L}\right) \\
\vdots & \ddots & \vdots \\
\mathbf{M}^{\left(m^{\prime},n^{\prime},k^{\prime}\right)}\left(\mathbf{r}_{L}-\mathbf{r}_{1}\right) & \ldots & \mathbf{M}^{\left(m^{\prime},n^{\prime},k^{\prime}\right)}\left(\mathbf{r}_{L}-\mathbf{r}_{L}\right)
\end{bmatrix}.
\end{align}
Similarly, the {matrices} ${\mathbf{E}_2}^{(m',n',k')} \in \mathbb{C}^{LJ \times QPJ}$ and ${\mathbf{E}_3}^{(m',n',k')} \in \mathbb{C}^{QPJ \times QPJ}$ consist of, respectively, $L Q$ block matrices of size $J\times PJ$ {and} $Q^2$ block matrices of size $PJ\times PJ$ as
\begin{align}
{\mathbf{E}_2}^{\left(m^{\prime}, n^{\prime},k^{\prime}\right)}=\begin{bmatrix}
\mathbf{N}^{\left(m^{\prime}, n^{\prime},k^{\prime}\right)} \left(\mathbf{r}_{1}- \mathbf{c}_{1}\right) & \ldots & \mathbf{N}^{\left(m^{\prime}, n^{\prime},k^{\prime}\right)}\left(\mathbf{r}_{1}- \mathbf{c}_{Q}\right) \\
\vdots & \ddots & \vdots \\
\mathbf{N}^{\left(m^{\prime}, n^{\prime},k^{\prime}\right)}\left(\mathbf{r}_{L}-\mathbf{c}_{1}\right) & \ldots & \mathbf{N}^{\left(m^{\prime}, n^{\prime},k^{\prime}\right)}\left(\mathbf{r}_{L}-\mathbf{c}_{Q}\right)
\end{bmatrix},
\end{align}
and
\begin{align}
{\mathbf{E}_3}^{\left(m^{\prime}, n^{\prime}\right)}=\begin{bmatrix}
\mathbf{P}^{\left(m^{\prime}, n^{\prime},k^{\prime}\right)} \left(\mathbf{c}_{1}-\mathbf{c}_{1}\right) & \ldots & \mathbf{P}^{\left(m^{\prime}, n^{\prime}\right)}\left(\mathbf{c}_{1}-\mathbf{c}_{Q}\right) \\
\vdots & \ddots & \vdots \\
\mathbf{P}^{\left(m^{\prime}, n^{\prime},k^{\prime}\right)}\left(\mathbf{c}_{Q}-\mathbf{c}_{1}\right) & \ldots & \mathbf{P}^{\left(m^{\prime}, n^{\prime},k^{\prime}\right)}\left(\mathbf{c}_{Q}-\mathbf{c}_{Q}\right)
\end{bmatrix}.
\end{align}

We rewrite the first conditions  of the dual certificate \eqref{eq:cert_1} and \eqref{eq:cert_2} as the following linear system of equations
as the following system of linear equations
\begin{align}
\underbrace{\left[\begin{array}{ccc}
{\mathbf{H}_1} & {\mathbf{H}_2}   \\
 {\mathbf{H}_2^H} & {\mathbf{H}_3}
\end{array}\right]}_{ = \mathbf{H} \in \mathbb{C}^{4(QPJ+LJ) \times 4(QPJ+LJ)}  }\left[\begin{array}{c}
\mathbf{j}_r \\
\mathbf{j}_c 
\end{array}\right]=\left[\begin{array}{c}
\mathbf{p}_r \\
\mathbf{p}_c
\end{array}\right],
\label{eq:system_coefficients}
\end{align}
where $\mathbf{p}_r$ and $\mathbf{p}_c$ {have been} defined in \eqref{eq:RHS_dual_certificate}. If the matrix $\mathbf{H}$ is invertible, the coefficients $\mathbf{j}_r$ ($\mathbf{j}_c$) for the radar (communications) polynomials could be obtained. To establish the invertibility of $\mathbf{H}$, we show that $\expec{\mathbf{H}}$ is invertible and then prove that $\mathbf{H}$ is close to $\expec{\mathbf{H}}$. Define the expectation 
\begin{align}
\overline{\mathbf{H}}=\mathbb{E}\left[\mathbf{H}\right]=\left[\begin{array}{ccc}
{\overline{\mathbf{H}}_1} & \overline{\mathbf{H}}_2    \\
 \overline{\mathbf{H}}_2^H   & \overline{{\mathbf{H}}}_3
\end{array}\right].
\label{eq:system_coefficients_expectation}
\end{align}
Note that $\overline{\mathbf{H}}_2$ is a zero-valued matrix of size $QPJ\times LJ$ because the expected value of the mixed terms that contain the product of independent random variables vanishes. Therefore, $\overline{\mathbf{H}}$ is a blockdiagonal matrix and its invertibility depends on the block matrices $\overline{\mathbf{H}}_1$ and $\overline{\mathbf{H}}_3$. Following  \cite[Proposition 2]{suliman2019exact}, $\overline{\mathbf{H}}_1$ and $\overline{\mathbf{H}}_3$ are invertible. 
The following Lemma \ref{lemma:distance H-E[H]} shows that $\mathbf{H}$ is very close to $\mathbb{E}\left[\mathbf{H}\right]$ and, hence, invertible with high probability. 
\begin{lemma}
\label{lemma:distance H-E[H]}
Consider the event 
$$
\mathcal{E}_{\epsilon_1}=\{   \|  \mathbf{H} - \mathbb{E}\left[\mathbf{H}\right] \| \leq \epsilon_1 \},
$$
for every real $\epsilon_1 \in (0,0.8)$. Then, the event $\mathcal{E}_{\epsilon_1}$ occurs with probability $1-4\delta$ for every $\delta>0$ provided that 
$$
MPN_R\geq\frac{172 \mu J}{\epsilon_1^2}\max\left(L,Q\right)\textnormal{log}\left(\max\left(\frac{8LJ}{\delta},\frac{8PQJ}{\delta}\right)\right).
$$
\label{lemma:concentration}
\end{lemma}
\begin{proof}
The proof follows from \cite[Appendix A]{vargas2023dual}, where $\mathbf{H}$ is expressed as the sum of zero-mean independent random matrices which are bounded using MBI.
\end{proof}

From Lemma \ref{lemma:distance H-E[H]}, and the inequalities \cite{suliman2019exact} $\|\mathbf{I}-\overline{\mathbf{H}}_1\|\leq 0.03254,$ 
and $\|\mathbf{I}-\overline{\mathbf{H}}_3\|\leq 0.03254,$ the matrix $\mathbf{H}$ is invertible with high probability. This satisfied the first two conditions of the dual certificate.  
\subsection{To satisfy conditions \eqref{eq:cert_3} and \eqref{eq:cert_4}}
To show that $\mathbf{f}_r$ and $\mathbf{f}_c$ achieves \eqref{eq:cert_3} and \eqref{eq:cert_4}, define the matrix 
\begin{align}
\boldsymbol{\Psi}_{r}^{(m^\prime,n^\prime,k^\prime)}(\mathbf{r})=& \frac{1}{\kappa^{m^\prime+n^\prime+k^\prime)}}\left[\begin{matrix}
\mathbf{M}^{(m^\prime,n^\prime,k^\prime)}\left(\mathbf{r}-\mathbf{r}_{1}\right)^{H} \\
\vdots \\
\mathbf{M}^{(m^\prime,n^\prime,k^\prime)}\left(\mathbf{r}-\mathbf{r}_{L}\right)^{H} \\ \frac{1}{\kappa}\mathbf{M}^{(m^{\prime}+1,n^\prime,k^\prime)}\left(\mathbf{r}-\mathbf{r}_{1}\right)^{H} \\
\vdots \\
\frac{1}{\kappa} \mathbf{M}^{(m^{\prime}+1,n^\prime,k^\prime)}\left(\mathbf{r}-\mathbf{r}_{L}\right)^{H}\\
\frac{1}{\kappa} \mathbf{M}^{(m^{\prime},n^{\prime}+1,k^\prime)}\left(\mathbf{r}-\mathbf{r}_{1}\right)^{H}\\
\vdots\\
\frac{1}{\kappa} \mathbf{M}^{(m^{\prime},n^{\prime}+1,k^\prime)}\left(\mathbf{r}-\mathbf{r}_{L}\right)^{H}\\
\frac{1}{\kappa} \mathbf{M}^{(m^{\prime},n^{\prime},k^\prime+1)}\left(\mathbf{r}-\mathbf{r}_{1}\right)^{H}\\
\vdots\\
\frac{1}{\kappa} \mathbf{M}^{(m^{\prime},n^{\prime},k^\prime+1)}\left(\mathbf{r}-\mathbf{r}_{L}\right)^{H}\\
\end{matrix}\right]
\end{align}
and its statistical expectation 
\begin{align}
\mathbb{E} &[\boldsymbol{\Psi}_{r}^{(m^\prime,n^\prime,k^\prime)}(\mathbf{r})] =\frac{1}{\kappa^{m^\prime+n^\prime+k^\prime}}\underbrace{\left[\begin{matrix}
K_{M}^{m^\prime}\left(\tau-[\bsym\tau_r]_{1}\right)^{*}K_{P}^{n^\prime}\left(\nu-[\bsym\nu_r]_{1}\right)^{*} K_{N_R}^{k^\prime}\left(\beta-[\bsym\beta_r]_{1}\right)^{*} \\\vdots\\K_{M}^{m^\prime}\left(\tau-[\bsym\tau_r]_{L}\right)^{*}K_{P}^{n^\prime}\left(\nu-[\bsym\nu_r]_{L}\right)^{*}K_{N_R}^{k^\prime}\left(\beta-[\bsym\beta_r]_{L}\right)^{*}  \\
\frac{1}{\kappa} K_{M}^{m^\prime+1}\left(\tau-[\bsym\tau]_{1}\right)^{*}K_{P}^{n^\prime}\left(\nu-[\bsym\nu]_{1}\right)^{*} K_{N_R}^{k^\prime}\left(\beta-[\bsym\beta_r]_{1}\right)^{*} \\
\vdots \\\frac{1}{\kappa} K_{M}^{m^\prime+1}\left(\tau-[\bsym\tau_r]_{L}\right)^{*}K_{P}^{n^\prime}\left(\nu-[\bsym\nu_r]_{L}\right)^{*} K_{N_R}^{k^\prime}\left(\beta-[\bsym\beta_r]_{L}\right)^{*}\\
\frac{1}{\kappa} K_{M}^{m^\prime}\left(\tau-[\bsym\tau]_{1}\right)^{*}K_{P}^{n^\prime+1}\left(\nu-[\bsym\nu_r]_{1}\right)^{*}K_{N_R}^{k^\prime}\left(\beta-[\bsym\beta_r]_{1}\right)^{*}\\\vdots\\\frac{1}{\kappa} K_{M}^{m^\prime}\left(\tau-[\bsym\tau]_{L}\right)^{*}K_{P}^{n^\prime+1}\left(\nu-[\bsym\nu]_{L}\right)^{*}K_{N_R}^{k^\prime}\left(\beta-[\bsym\beta_r]_{L}\right)^{*}\\\frac{1}{\kappa} K_{M}^{m^\prime}\left(\tau-[\bsym\tau]_{1}\right)^{*}K_{P}^{n^\prime}\left(\nu-[\bsym\nu_r]_{1}\right)^{*}K_{N_R}^{k^\prime+1}\left(\beta-[\bsym\beta_r]_{1}\right)^{*}\\\vdots\\\frac{1}{\kappa} K_{M}^{m^\prime}\left(\tau-[\bsym\tau]_{L}\right)^{*}K_{P}^{n^\prime}\left(\nu-[\bsym\nu]_{L}\right)^{*}K_{N_R}^{k^\prime+1}\left(\beta-[\bsym\beta_r]_{L}\right)^{*}\\
\end{matrix}\right] }_{ \triangleq\bsym{\psi}_r^{(m^\prime,n^\prime,k^\prime)}(\mathbf{r})}\otimes \mathbf{I}_{J}\nonumber
\end{align}

The corresponding communciations matrix $\bsym{\Psi}_c^{(m^\prime,n^\prime,k^\prime)}$ is built similarly by replacing the kernel matrices $\mathbf{M}^{m^\prime,n^\prime,k^\prime}(\mathbf{r})$, with $\mathbf{N}^{m^\prime,n^\prime,k^\prime}(\mathbf{c})$. Its expected value, \textit{mutatis mutandis}, is $\mathbb{E}[\bsym{\Psi}_c^{(m^\prime,n^\prime,k^\prime)}(\mathbf{r})]=\frac{1}{\kappa^{m^\prime+n^\prime+k^\prime}}\bsym{\psi}_r^{(m^\prime,n^\prime,k^\prime)}\otimes \mathbf{0}_{PJ\times J} = \mathbf{0}_{4JPL\times J}$ because $\mathbb{E}[\mathbf{d}_{\widetilde{n}}\mathbf{b}_n]=\mathbf{0}_{JP\times J}$

To simplify the bound computation, we follow the procedure outlined in \cite{tang2013compressed}. Express $\mathbf{H}^{-1}$ in terms of the matrices $\mathbf{L}_1 \in \mathbb{C}^{4LJ\times LJ}, \mathbf{R}_1 \in \mathbb{C}^{4LJ\times 3JL}\widehat{\mathbf{L}}_1 \in \mathbb{C}^{4QPJ\times LJ},\widehat{\mathbf{R}}_1 \in \mathbb{C}^{4LJ\times 3LJ}, \mathbf{L}_2 \in \mathbb{C}^{4LJ\times PJQ}, \mathbf{R}_2 \in \mathbb{C}^{4LJ\times 3PJQ}, \widehat{\mathbf{L}}_2 \in \mathbb{C}^{4QPJ\times PJQ},\widehat{\mathbf{R}}_2 \in \mathbb{C}^{4QPJ\times 3PJQ},\widehat{\mathbf{L}}_2 \in \mathbb{C}^{4QPJ\times PQJ},\widehat{\mathbf{R}}_2 \in \mathbb{C}^{4LJ\times 3PQJ}$ as 
$$\mathbf{H}^{-1} \triangleq \left[\begin{array}{cccc}
\mathbf{L}_1 & \mathbf{R_1} & \mathbf{L}_2 & \mathbf{R_2}\\\widehat{\mathbf{L}}_1 & \widehat{\mathbf{R}}_1 & \widehat{\mathbf{L}}_2 & \widehat{\mathbf{R}}_2
\end{array}\right].
$$
Consequently, 
$$
\overline{\mathbf{H}}^{-1} = \begin{bmatrix}
\mathbf{L}_1^\prime \otimes \mathbf{I}_J& \mathbf{R}_1^\prime \otimes \mathbf{I}_J& \mathbf{0}_{4LJ\times PJQ} & \mathbf{0}_{4LJ\times 3PJQ} \\\mathbf{0}_{4QPJ\times LJ} & \mathbf{0}_{4QPJ\times 3LJ} & \widehat{\mathbf{L}}_2^\prime\otimes \mathbf{I}_{PJ}& \widehat{\mathbf{R}}_2^\prime\otimes \mathbf{I}_{PJ}
\end{bmatrix}.
$$ 
Then, the solutions of \eqref{eq:system_coefficients} are
$\mathbf{j}_r=\mathbf{L}_1\widetilde{\mathbf{u}}+\mathbf{L}_2\widetilde{\mathbf{v}}$
and$\mathbf{j}_c=\widehat{\mathbf{L}}_1\widetilde{\mathbf{u}}+\widehat{\mathbf{L}}_2\widetilde{\mathbf{v}}$
where $\widetilde{\mathbf{u}}=\textrm{sign}(\bsym{\alpha}_r)\otimes\mathbf{u}$ and $\widetilde{\mathbf{v}}=\textrm{sign}(\bsym{\alpha}_c)\otimes\mathbf{v}$.

Using the expressions above, the radar polynomial in \eqref{eq:polyr_random_v1}  becomes 
\begin{align}
&\frac{1}{\kappa^{{m^\prime+n^\prime+k^\prime}}} \mathbf{f}^{(m^\prime,n^\prime,k^\prime)}_r(\mathbf{r})  \nonumber\\&= \begin{bmatrix}
     \bsym{\Psi}_r^{(m^\prime,n^\prime,k^\prime)}(\mathbf{r})^H  & \bsym{\Psi}_c^{(m^\prime,n^\prime,k^\prime)}(\mathbf{r})^H
\end{bmatrix}\begin{bmatrix}   \mathbf{j}_r \nonumber\\
     \mathbf{j}_c 
\end{bmatrix}\\
&= \underbrace{\bsym{\Psi}_r^{(m^\prime,n^\prime,k^\prime)}\mathbf{r})^H \mathbf{L}_1\widetilde{\mathbf{u}}}_{=\bsym\Theta_r^{(m^\prime,n^\prime,k^\prime)}(\mathbf{r})}+\underbrace{ \bsym{\Psi}_r^{(m^\prime,n^\prime,k^\prime)}\mathbf{L}_2\widetilde{\mathbf{v}}}_{=\widetilde{\boldsymbol{\Theta}}_r^{(m^\prime,n^\prime,k^\prime)}(\mathbf{r})}+\underbrace{\bsym{\Psi}_c^{(m^\prime,n^\prime,k^\prime)}(\mathbf{r})^H  \widehat{\mathbf{L}}_1\widetilde{\mathbf{u}}}_{=\boldsymbol{\Theta}_c^{(m^\prime,n^\prime,k^\prime)}(\mathbf{r})}+\underbrace{\bsym{\Psi}_c^{(m^\prime,n^\prime,k^\prime)}(\mathbf{r})^H\widehat{\mathbf{L}}_2\widetilde{\mathbf{v}}}_{= \widetilde{\boldsymbol{\Theta}}_c^{(m^\prime,n^\prime,k^\prime)}(\mathbf{r})}.\label{eq:fr_non_decom}
\end{align}

The matrix $\bsym{\Theta}^{(m^\prime,n^\prime,k^\prime)}_r(\mathbf{r})$ is 
\begin{align}
&\boldsymbol{\Theta}_r^{(m^\prime,n^\prime,k^\prime)}(\mathbf{r})=\mathbb{E}\left[ \bsym{\Psi}_r^{(m^\prime,n^\prime,k^\prime)}(\mathbf{r})\right]^{H}\left(\mathbf{L}_1^{\prime} \otimes \mathbf{I}_{J}\right) \widetilde{\mathbf{u}}+\underbrace{\left(\bsym{\Psi}_r^{(m^\prime,n^\prime,k^\prime)}(\mathbf{r})-\mathbb{E} \left[\bsym{\Psi}_r^{(m^\prime,n^\prime,k^\prime)}(\mathbf{r})\right]\right)^{H} \mathbf{L}_1 \widetilde{\mathbf{u}}}_{=\mathbf{J}_{1}^{(m^\prime,n^\prime,k^\prime)}(\mathbf{r})} +\nonumber\\&\underbrace{\mathbb{E}\left[ \bsym{\Psi}_r^{(m^\prime,n^\prime,k^\prime)}(\mathbf{r})\right]^{H}\left(\mathbf{L}_1-\mathbf{L}_1^{\prime} \otimes \mathbf{I}_{J}\right) \widetilde{\mathbf{u}}}_{=\mathbf{J}_{2}^{(m^\prime,n^\prime,k^\prime)}(\mathbf{r})}.\label{eq:decomposition_theta_r}
\end{align}
We rewrite its first term as
$$
\begin{aligned}
&\mathbb{E}\left[ \bsym{\Psi}_r^{(m^\prime,n^\prime,k^\prime)}(\mathbf{r})\right]^{H}\left(\mathbf{L}_1^{\prime} \otimes \mathbf{I}_{J}\right) \widetilde{\mathbf{u}}\triangleq \frac{1}{\kappa^{m'+n'+,k'}}f^{(m',n',k')}(\mathbf{r})\mathbf{u},
\end{aligned}
$$
where
the scalar-valued polynomial 
\begin{align}
    f^{(m^\prime, n^\prime,k^\prime)} &= \sum_{\ell=1}^{L} K^{(m^\prime, n^\prime,k^\prime)}(\mathbf{r-r}_\ell)\overline{\bsym{\beta}}_\ell + K^{(m^\prime+1, n^\prime,k^\prime)}(\mathbf{r-r}_\ell)\overline{\bsym{\gamma}}_\ell  \nonumber\\
    & + K^{(m^\prime, n^\prime+1,,k^\prime)}(\mathbf{r-r}_\ell)\overline{\bsym{\zeta}}_\ell+K^{(m^\prime, n^\prime,k^\prime+1)}(\mathbf{r-r}_\ell)\overline{\bsym{\varpi}}_\ell.
\end{align}
For the second term of \eqref{eq:fr_non_decom}, define
$$
\begin{aligned}
\widetilde{\boldsymbol{\Theta}}_r^{(m^\prime,n^\prime,k^\prime)}(\mathbf{r})=&\underbrace{\left(\bsym{\Psi}_r^{(m^\prime,n^\prime,k^\prime)}(\mathbf{r})^H-\mathbb{E} \bsym{\Psi}_r^{(m^\prime,n^\prime,k^\prime)}(\mathbf{r})^{H}\right) \mathbf{L}_2 \widetilde{\mathbf{v}}}_{\mathbf{J}_{3}^{(m^\prime,n^\prime,k^\prime)}(\mathbf{r})}+ \underbrace{\mathbb{E}\left[ \bsym{\Psi}_r^{(m^\prime,n^\prime,k^\prime)}(\mathbf{r})\right]^{H} \left(\mathbf{L}_2 \right) \widetilde{\mathbf{v}}(\mathbf{r})}_{\mathbf{J}_{4}^{(m^\prime,n^\prime,k^\prime)}(\mathbf{r})},
\end{aligned} 
$$
Finally, substituting various expressions above, \eqref{eq:fr_non_decom}  becomes
\begin{align}
      \frac{1}{\kappa^{m'+n'+k'}}\mathbf{f}_r^{(m^\prime, n^\prime,k^\prime)}(\mathbf{r}) =& \frac{1}{\kappa^{m'+n'+k'}}   f^{(m^\prime, n^\prime,k^\prime)}(\mathbf{r})\mathbf{u} +  \mathbf{J}_1^{(m^\prime,n^\prime,k^\prime)}(\mathbf{r}) +\mathbf{J}_2^{(m^\prime,n^\prime,k^\prime)}(\mathbf{r}) \nonumber\\
      & + \mathbf{J}_3^{(m^\prime,n^\prime,k^\prime)}(\mathbf{r}) +\mathbf{J}_4^{(m^\prime,n^\prime,k^\prime)}(\mathbf{r}) + \boldsymbol{\Theta}_c^{(m^\prime,n^\prime,k^\prime)}(\mathbf{r}) +\widetilde{\boldsymbol{\Theta}}_c^{(m^\prime,n^\prime,k^\prime)}(\mathbf{r}).
      \label{decomp_f}
\end{align}

Our goal is to demonstrate that $ \frac{1}{\kappa}\mathbf{f}_r^{(m^\prime, n^\prime,k^\prime)}(\mathbf{r}) $ is close to $\frac{1}{\kappa^{m'+n'+k^\prime}}   f^{(m^\prime, n^\prime,k^\prime)}(\mathbf{r})\mathbf{u}$ and, similarly, $\frac{1}{\kappa}\mathbf{f}_c^{(m^\prime, n^\prime,k^\prime)}(\mathbf{c}) $ is close to $\frac{1}{\kappa^{m'+n'+k^\prime}}   f^{(m^\prime, n^\prime,k^\prime)}(\mathbf{c})\mathbf{v}$. We start by proving that the spectral norms of ~$\mathbf{J}_1^{(m^\prime,n^\prime,k^\prime)}(\mathbf{r})$, $\mathbf{J}_2^{(m^\prime,n^\prime,k^\prime)}(\mathbf{r})$, $\mathbf{J}_3^{(m^\prime,n^\prime,k^\prime)}(\mathbf{r}),\mathbf{J}_4^{(m^\prime,n^\prime,k^\prime)}(\mathbf{r})$, $ \boldsymbol{\Theta}_c^{(m^\prime,n^\prime,k^\prime)}(\mathbf{r})$, and  ~$\widetilde{\boldsymbol{\Theta}}_c^{(m^\prime,n^\prime,k^\prime)}(\mathbf{r})$ are small with high probability on the set of grid points $\Omega_{\textrm{grid}}$. 
Then, we show that $ \frac{1}{\kappa^{m'+n'+k'}}\mathbf{f}_r^{(m^\prime, n^\prime,k^\prime)}(\mathbf{r}) $ is close to $\frac{1}{\kappa^{m'+n'+k'}}   f^{(m^\prime, n^\prime,k^\prime)}(\mathbf{r})\mathbf{u}$ on $\Omega_{\textrm{grid}}$. Finally, we show that the polynomial constructed as above satisfies the expected condition $\|\mathbf{f}_r(\mathbf{r})\|_2<1$, $\mathbf{r}\setminus \mathcal{R}$. With a proper grid size, we extend this result for the continuous domain $[0,1]^3$. 
\subsubsection{Bounds {for} $\Vert\mathbf{J}_i^{(m^\prime,n^\prime,k^\prime)}(\mathbf{r})\Vert_2$,$\Vert\boldsymbol{\Theta}_c^{(m^\prime,n^\prime,k^\prime)}(\mathbf{r})\Vert_2$, and  ~$\Vert\widetilde{\boldsymbol{\Theta}}_c^{(m^\prime,n^\prime,k^\prime)}(\mathbf{r})\Vert_2$} 
\label{section:bound_J1} 
We use the fact that ${\boldsymbol{\Theta}}_c^{(m^\prime,n^\prime,k^\prime)}(\mathbf{r})$ and $\Vert\widetilde{\boldsymbol{\Theta}}_c^{(m^\prime,n^\prime,k^\prime)}(\mathbf{r})\Vert_2$ can be expressed as the  sum of random zero-mean matrices, and consequently, matrix Bernstein inequality is applicable. In the sequel, we state these bounds in Lemmata \ref{lemma:J1_last}, \ref{lemma:J2_last}, 
 \ref{lemma:J3_last}, \ref{lemma:J4_last}, \ref{lemma:theta_last}, and \ref{lemma:theta_tilde_last}, whose proofs follow \textit{ceteris paribus} from \cite[Lemata 4, 5, 6, 7, 8, and 9]{vargas2023dual}, respectively. They are based on decomposing the matrices in the sum of zero-mean random matrices and applying MBI. Then using the union bound for all the derivatives concludes the proof for each Lemma.

\begin{lemma}\label{lemma:J1_last}
{Assume} that $\mathbf{u} \in \mathbb{C}^{J}$ {is} sampled {on} the complex unit sphere. Let $0 < \delta < 1 $ the failure probability, and $\epsilon_2>0$. There exists a numerical constant $C$ such that if 
\begin{align*}
    MPN_R \geq &C\mu \operatorname{max}(L,Q)J\operatorname{max}\left\{\frac{1}{\epsilon_2^2}\log\left(\frac{\vert\Omega_{\textnormal{\textrm{grid}}}\vert JL}{\delta}\right)\right.\log^2\left(\frac{\vert\Omega_{\textnormal{\textrm{grid}}}\vert J}{\delta}\right),\left.\log\left(\frac{LJ}{\delta}\right),\log\left(\frac{PQJ}{\delta}\right)\right\},
\end{align*}
the following inequality is satisfied 
\begin{align*}
    \mathbb{P}\left\{\underset{\widetilde{\mathbf{r}}_d\in\Omega_{\textnormal{\text{grid}}}}{sup}\left\Vert\mathbf{J}_1^{(m^\prime,n^\prime,k^\prime)})(\widetilde{\mathbf{r}})\right\Vert_2\geq   \epsilon_2, m^\prime,n^\prime,k^\prime=0,1,2,3 \right\}\leq 64\delta.
\end{align*}
\end{lemma}

\begin{lemma}\label{lemma:J2_last}
{Assume} that $\mathbf{u} \in \mathbb{C}^{J}$ {is sampled on} the complex unit sphere. Let $0 < \delta < 1 $ the failure probability, and $\epsilon_3>0$. There exists a numerical constant $C$ such that if
\begin{align*}
    MPN_R \geq &\frac{C\mu J\operatorname{max}(L,Q)}{\epsilon_3^2}\log^2\left(\frac{\vert\Omega_{\textnormal{\textrm{grid}}}\vert J}{\delta}\right)\operatorname{max}\left\{\log\left(\frac{JL}{\delta}\right),\log\left(\frac{PQJ}{\delta}\right)\right\},
\end{align*}
{the following inequality is satisfied}
\begin{align*}
    \mathbb{P}\left\{\underset{\widetilde{\mathbf{r}}_d\in\Omega_{\text{grid}}}{sup}\left\Vert\mathbf{J}_2^{(m^\prime,n^\prime,k^\prime)}(\widetilde{\mathbf{r}})\right\Vert_2\geq   \epsilon_3, m^\prime, n^\prime,k^\prime=0,1,2,3\right\}\leq 48\delta.
\end{align*}
\end{lemma}
\begin{lemma}\label{lemma:J3_last}
{Assume} that $\mathbf{v} \in \mathbb{C}^{J}$ {is sampled on} the complex unit sphere. Let $0 < \delta < 1 $ the failure probability and $\epsilon_4>0$. There exists a numerical constant $C$ such that if
\begin{align*}
    MPN_R \geq & C\mu \operatorname{max}(L,Q)J\operatorname{max}\left\{\frac{1}{\epsilon_4^2}\log\left(\frac{\vert\Omega_{\textnormal{\textrm{grid}}}\vert JL}{\delta}\right)\right.\log^2\left(\frac{\vert\Omega_{\textnormal{\textrm{grid}}}\vert J}{\delta}\right),\left.\log\left(\frac{JL}{\delta}\right),\log\left(\frac{PQJ}{\delta}\right)\right\},
\end{align*}
{the following inequality is satisfied}
\begin{align*}
    \mathbb{P}\left\{\underset{\widetilde{\mathbf{r}}_d\in\Omega_{\textnormal{\text{grid}}}}{sup}\left\Vert\mathbf{J}_3^{(m^\prime,n^\prime,k^\prime)}\right(\widetilde{\mathbf{r}})\Vert_2\geq   \epsilon_4, m^\prime,n^\prime,k^\prime=0,1,2,3 \right\}\leq 64\delta.
\end{align*}
\end{lemma}
\begin{lemma}\label{lemma:J4_last}
{Assume} that $\mathbf{v} \in \mathbb{C}^{J}$ {is sampled on} the complex unit sphere. Let $0<\delta<1$ be the failure probability and $\epsilon_5>0$ a constant. There exists a numerical constant $C$ such that if
\begin{align*}
    MPN_R \geq& C\mu \operatorname{max}(L,Q)J\\
    &\operatorname{max}\left\{\log^2\left(\frac{\vert\Omega_{\textnormal{\textrm{grid}}}\vert J}{\delta}\right),\log\left(\frac{JL}{\delta}\right),\log\left(\frac{PQJ}{\delta}\right)\right\},
\end{align*}
{the following inequality is satisfied}
\begin{align*}
    \mathbb{P}\left\{\underset{\widetilde{\mathbf{r}}_d\in\Omega_{\textnormal{\textrm{grid}}}}{sup}\left\Vert\mathbf{J}_4^{(m^\prime,n^\prime,k^\prime)}(\widetilde{\mathbf{r}})\right\Vert_2\geq   \epsilon_5, m^\prime, n^\prime, k^\prime=0,1,2,3 \right\}\leq 48\delta.
\end{align*} 
\end{lemma}

\begin{lemma}\label{lemma:theta_last}
{Assume} that $\mathbf{v} \in \mathbb{C}^{PJ}$ {is sampled on} the complex unit sphere. Let $0<\delta<1$ be the failure probability and $\epsilon_6>0$ a constant. There exists a numerical constant $C$ such that if
\begin{align*}
    MPN_R \geq & C\mu \operatorname{max}(L,Q)J\operatorname{max}\left\{\frac{1}{\epsilon_6^2}\log\left(\frac{\vert\Omega_{\textnormal{\textrm{grid}}}\vert PQJ}{\delta}\right)\right.\log^2\left(\frac{\vert\Omega_{\textnormal{\textrm{grid}}}\vert J}{\delta}\right),\left.\log\left(\frac{LJ}{\delta}\right),\log\left(\frac{PQJ}{\delta}\right)\right\},
\end{align*}
{then, the following inequality is satisfied}
\begin{align*}
    \mathbb{P}\left\{\underset{\widetilde{\mathbf{r}}_d\in\Omega_{\textnormal{\textrm{grid}}}}{sup}\left\Vert{\bsym{\Theta}}_c^{(m^\prime,n^\prime,k^\prime)}(\widetilde{\mathbf{r}})\right\Vert_2\geq   \epsilon_6, m^\prime, n^\prime,k^\prime=0,1,2,3 \right\}\leq 64\delta.
\end{align*}
\end{lemma}

\begin{lemma}\label{lemma:theta_tilde_last}
{Assume} that $\mathbf{v} \in \mathbb{C}^{PJ}$ {is sampled on} the complex unit sphere. Let $0<\delta<1$ be the failure probability and $\epsilon_6>0$ a constant. There exists a numerical constant $C$ such that if
\begin{align*}
    MPN_R \geq& C\mu \operatorname{max}(L,Q)J\operatorname{max}\left\{\log^2\left(\frac{\vert\Omega_{\textnormal{\textrm{grid}}}\vert J}{\delta}\right),\log\left(\frac{LJ}{\delta}\right),\log\left(\frac{PQJ}{\delta}\right)\right\},
\end{align*}
{the following inequality is satisfied}
\begin{align*}
    \mathbb{P}\left\{\underset{\widetilde{\mathbf{r}}_d\in\Omega_{\textnormal{\textrm{grid}}}}{sup}\left\Vert\widetilde{\bsym{\Theta}}_r^{(m^\prime,n^\prime,k^\prime)}(\widetilde{\mathbf{r}})\right\Vert_2\geq   \epsilon_7, m^\prime, n^\prime,k^\prime=0,1,2,3 \right\}\leq 48\delta.
\end{align*}
\end{lemma}

\subsubsection{$ \frac{1}{\kappa}\mathbf{f}_r^{(m^\prime, n^\prime)} $ is close to $\frac{1}{\kappa^{m'+n'}}   f^{(m^\prime, n^\prime)}\mathbf{u}$ on $\Omega_{\textnormal{\textrm{grid}}}$}
We obtain this as a consequence of the following Lemma~\ref{lemma:bound_f_f_tilde_discrete}.
\begin{lemma}\label{lemma:bound_f_f_tilde_discrete}
Assume that $\Omega_{\textrm{grid}} \in [0,1)^2$ is a finite set of points. Let $0<\delta<1$ be the failure probability and $\epsilon>0$ a constant. Then, there exists a numerical constant $C$ such that if
\begin{align*}
    &MP \geq  C\mu \operatorname{max}(L,Q)J\operatorname{max}\left\{\frac{1}{\epsilon^2}\log\left(\frac{\vert\Omega_{\textnormal{\textrm{grid}}}\vert PQJ}{\delta}\right),\right.\\&\frac{1}{\epsilon^2}\log\left(\frac{\vert\Omega_{\textnormal{\textrm{grid}}}\vert LJ}{\delta}\right),\log^2\left(\frac{\vert\Omega_{\textnormal{\textrm{grid}}}\vert J}{\delta}\right),\left.\log\left(\frac{LJ}{\delta}\right),\log\left(\frac{PQJ}{\delta}\right)\right\},
\end{align*}
then, for the event 
\begin{align*}
 \mathcal{E}_{\epsilon}=\left\{\underset{\widetilde{\mathbf{r}}_d\in\Omega_{\textnormal{\textrm{grid}}}}{\sup}\left\|\frac{1}{\kappa^{m'+n'}}\mathbf{f}_r^{(m^\prime, n^\prime)}(\widetilde{\mathbf{r}}) -\frac{1}{\kappa^{m'+n'}}   f^{(m^\prime, n^\prime)}(\widetilde{\mathbf{r}})\mathbf{u}\right\|_2\leq \frac{\epsilon}{3},\right.
\left.m^\prime,n'=0,1,2,3\right\},
\end{align*}
we have
$$
\mathbb{P}[\mathcal{E}_{\epsilon}]\geq 1 - \delta.
$$
\end{lemma}
\begin{proof}
Using Lemmata \ref{lemma:J1_last}-
\ref{lemma:theta_tilde_last}, it follows that the event $\mathcal{E}_{\epsilon}$ holds with a high probability. This completes the proof. 
\end{proof}

\subsubsection{$\|\mathbf{f}_r(\mathbf{r})\|_2<1$ for $\mathbf{r} \in [0,1]^3 \setminus \mathcal{R}$}
Define the minimum separation condition for the off-the-grid delay, Doppler, and DoA parameters as, respectively,
$ \min\left( \left\vert [\bsym{\tau
    }_r]_{\ell} - [\bsym\tau]_r]_{\ell^\prime}\right\vert,\left\vert[\bsym{\tau
    }_c]_{q} - [\bsym\tau]_c]_{q^\prime}\right\vert\right) \triangleq \Delta_\tau,
    \min\left( \left\vert [\bsym{\nu
    }_r]_{\ell} - [\bsym\nu]_r]_{\ell^\prime}\right\vert,\left\vert[\bsym\nu_c]_{q} - [\bsym\nu_c]_{q^\prime}\right\vert\right)\geq \Delta_\nu$, $\min( \vert [\bsym{\beta
    }_r]_{\ell}$ $ - [\bsym\beta]_r]_{\ell^\prime}\vert,\left\vert[\bsym\beta_c]_{q} - [\bsym\beta_c]_{q^\prime}\right\vert\geq \Delta_\beta$, 
where $\ell \neq\ell^\prime, q \neq q^{\prime}$, and $\vert a-b\vert$ is the wrap-around metric on [0,1), e.g. $|0.2-0.1| = 0.1$ but $|0.8-0.1|=0.3$. The following lemma shows that $ \frac{1}{\kappa}\mathbf{f}_r^{(m^\prime, n^\prime, k^\prime)} $ is close to $\frac{1}{\kappa^{m'+n'+k'}}   f^{(m^\prime, n^\prime, k^\prime)}\mathbf{u}$ everywhere on $[0,1)^3$.

\begin{lemma}\label{lemma:certificate_less_1_everywhere}
Assume that $\Delta_\tau \leq 1/M$ and $\Delta_\nu \leq 1/P$. Denote the failure probability by $0<\delta<1$. Then, for all $\mathbf{r}=[\tau,\nu,\beta]$, $m^\prime,n',k'=0,1,2,3$, and $\epsilon>0$, there exists a numerical constant C such that if 
\begin{align*}
    &MPN_R \geq  C\mu \operatorname{max}(L,Q)J\operatorname{max}\left\{\frac{1}{\epsilon^2}\log\left(\frac{MPN_R QJ}{\delta}\right),\right.\\
    &\hspace{-2mm}\frac{1}{\epsilon^2}\log\left(\frac{MPN_R LJ}{\delta}\right),\log^2\left(\frac{MPN_R J}{\delta}\right),\left.\log\left(\frac{LJ}{\delta}\right),\log\left(\frac{PQJ}{\delta}\right)\right\},
\end{align*}
the following inequality is satisfied
$$
\left\|\frac{1}{\kappa^{m'+n',k'}}\left(\mathbf{f}_r^{(m^\prime, n^\prime,k')}(\mathbf{r}) -   f^{(m^\prime, n^\prime,k')}(\mathbf{r})\mathbf{u}\right)\right\|_2\leq \frac{\epsilon}{3}.
$$
\end{lemma}
\begin{proof}
The proof follows from \cite[Appendix 
  E]{vargas2023dual} except that we select the proper grid size as $\left|\Omega_{grid}\right| = \frac{4C(MPN_R)^2}{\epsilon}$. Then,  conditioned on the event $\mathcal{E}_{\varepsilon_1}$ in Lemma 11, it follows that for the selected grid, $\mathbf{f}_r^{(m^\prime, n^\prime,k')}(\mathbf{r}) $ and $  f^{(m^\prime, n^\prime,k')}(\mathbf{r})\mathbf{u}$ are close with high probability.
\end{proof}

The following lemma states that $\|\mathbf{f}_r(\mathbf{r})\|_2<1$ with a high probability.
\begin{lemma}\label{lemma:certificate_less_1}
There exists a constant $C$ such that if 
\begin{align}
\label{eq:sample_complexity}
    &MPN_R \geq  C\mu \operatorname{max}(L,Q)J\log^2\left(\frac{MPN_R J}{\delta}\right)\operatorname{max}\left\{\log\left(\frac{MPN_RQJ}{\delta}\right),\log\left(\frac{MPN_RLJ}{\delta}\right)\right\},
\end{align}
then with probability $1-\delta$, where $0<\delta<1$, $\|\mathbf{f}_r(\mathbf{r})\|_2<1$, for all $\mathbf{r} \in [0,1]^3 \setminus \mathcal{R}$.
\end{lemma}
\begin{proof}
The proof requires showing $\|\mathbf{f}_r(\mathbf{r})\|_2<1$ in the set of points near and far from $\mathbf{r}_j \in \mathcal{R}$. We refer the reader to \cite[Lemma 14]{suliman2021mathematical} for details, wherein an analogous result for multi-dimensional bivariate dual polynomial in a single-BD problem has been proved. In our DBD setting, the dual polynomials are similarly multi-dimensional bi-variate but contain cross-terms (see (105)). Following the procedure in \cite[Lemma 14]{suliman2021mathematical} but instead using Lemma~\ref{lemma:certificate_less_1_everywhere} for our specific polynomial with $\epsilon=0.0005$ concludes the proof. 
\end{proof}

\begin{figure*}[!t]
    \centering
    \includegraphics[width=\linewidth]{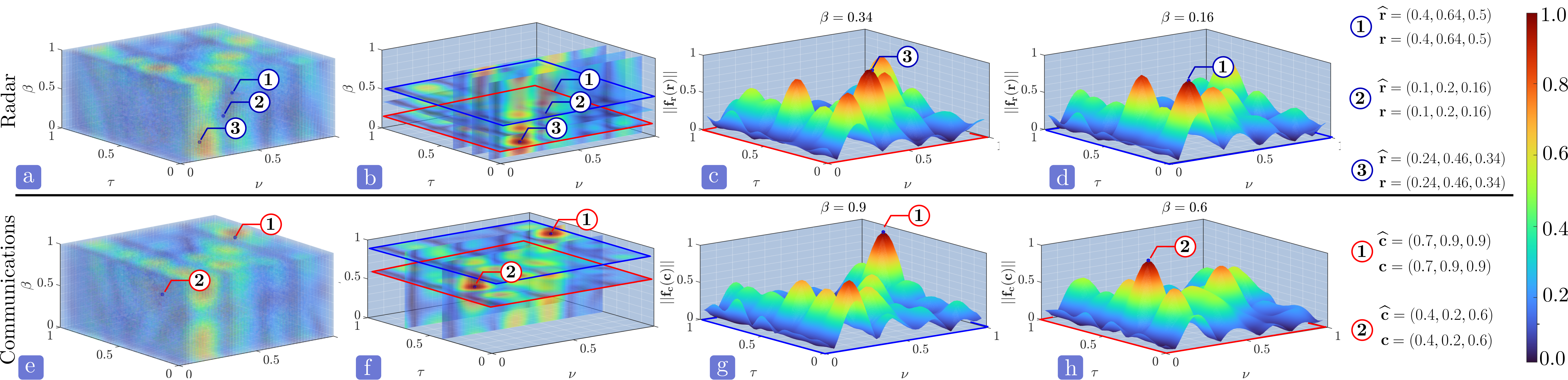}
    \caption{Channel parameter localization in the 3-D dual polynomials. The parameter estimates are localized in the 3-D delay-Doppler-DoA hyperplane when the polynomials' norm is unity. Here, we set $M = 13$, $P=9, N_R = 5$,  $L=4$, and $J=3$. Top panel: (a) Scatter plot of 3-D $\mathbf{f}_r(\mathbf{r})$ polynomial with the color indicating the value of its norm evaluated at ${\tau, \nu, \beta}$ values. (b) As in (a), but for only 4 values of $\beta$ that correspond to the estimated DoAs of $L=4$ targets. These are the norm values of $\mathbf{f}_r(\mathbf{r})$ in the delay-Doppler plane for 4 fixed $\beta$ values. The rectangles with blue and red borders indicate the specific $\beta$ values of $0.14$ and $0.36$, for which the 2-D plot of $\Vert \mathbf{f}_r(\mathbf{r})\Vert_2^2 $ is shown in (c) and (d), respectively. Bottom panel: (e) As in (a), but for $\mathbf{f}_c(\mathbf{c})$ with $Q=2$ communications propagation paths. (f) As in (b), but for $\Vert \mathbf{f}_c(\mathbf{c})\Vert_2^2 $ with $Q=2$ delay-Doppler slices. The blue and red border rectangles indicate the specific $\beta$ values of $0.5$ and $0$, for which the 2-D plot of $\Vert \mathbf{f}_c(\mathbf{c})\Vert_2^2 $ is shown in (g) and (h), respectively. }
    \label{fig:results_dual}
\end{figure*}
\section{Numerical Experiments}
\label{sec:experiments}

\begin{figure*}[!t]
    \centering
    \includegraphics[width=\linewidth]{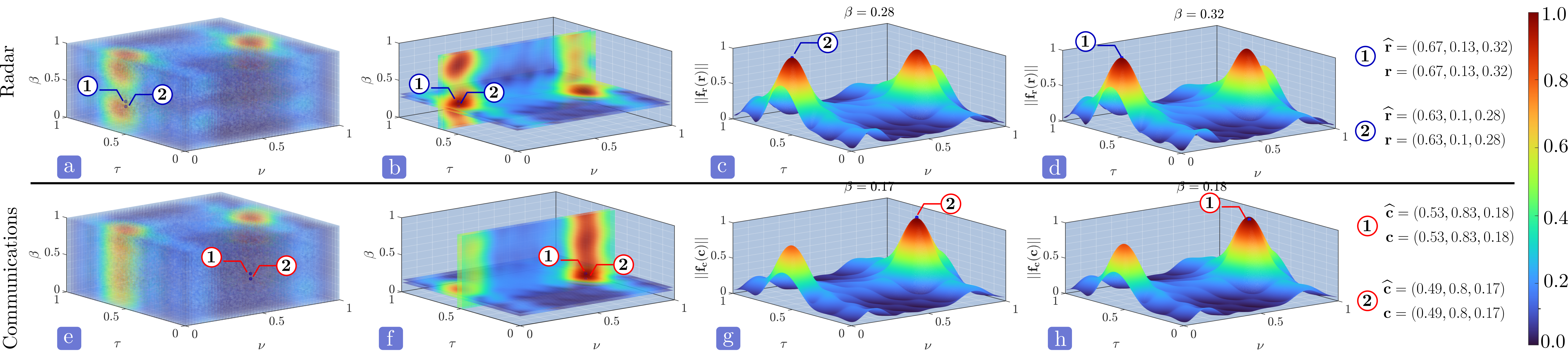}
    \caption{As in Fig. \ref{fig:results_dual}, but for closely spaced channel parameters. Here, $M =17$ time samples, $P=11 $ pulses/messages, $N_R=9$ number of receiver antennas and subspace size $J=3$. 
    }
    \label{fig:results_closely_parameters}
\end{figure*}
\begin{figure}
    \centering
\includegraphics[width=1.0\columnwidth]{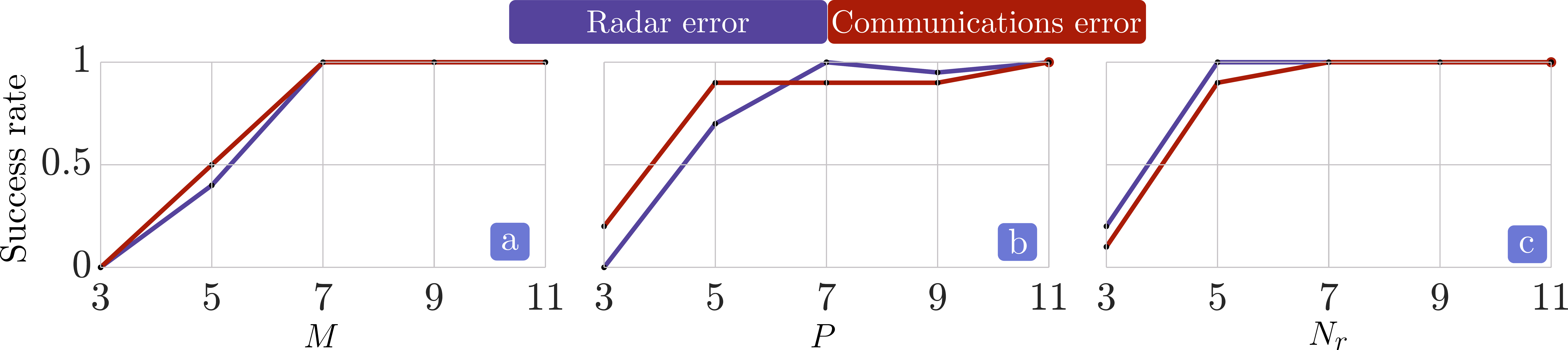}\qquad
    \caption{Success rate with respect to the (a) number of samples $M$ {for fixed $P= 5, N_R = 5$ }, (b) number of pulses/messages $P$ {for fixed $M= 5, N_r = 5$ }, and (c) number of antennas $N_R$ {for fixed $M= 5, P = 5$ }. 
    }
    \label{fig:stats}
\end{figure}
\begin{figure}
    \centering
    \includegraphics[width=0.5\linewidth]{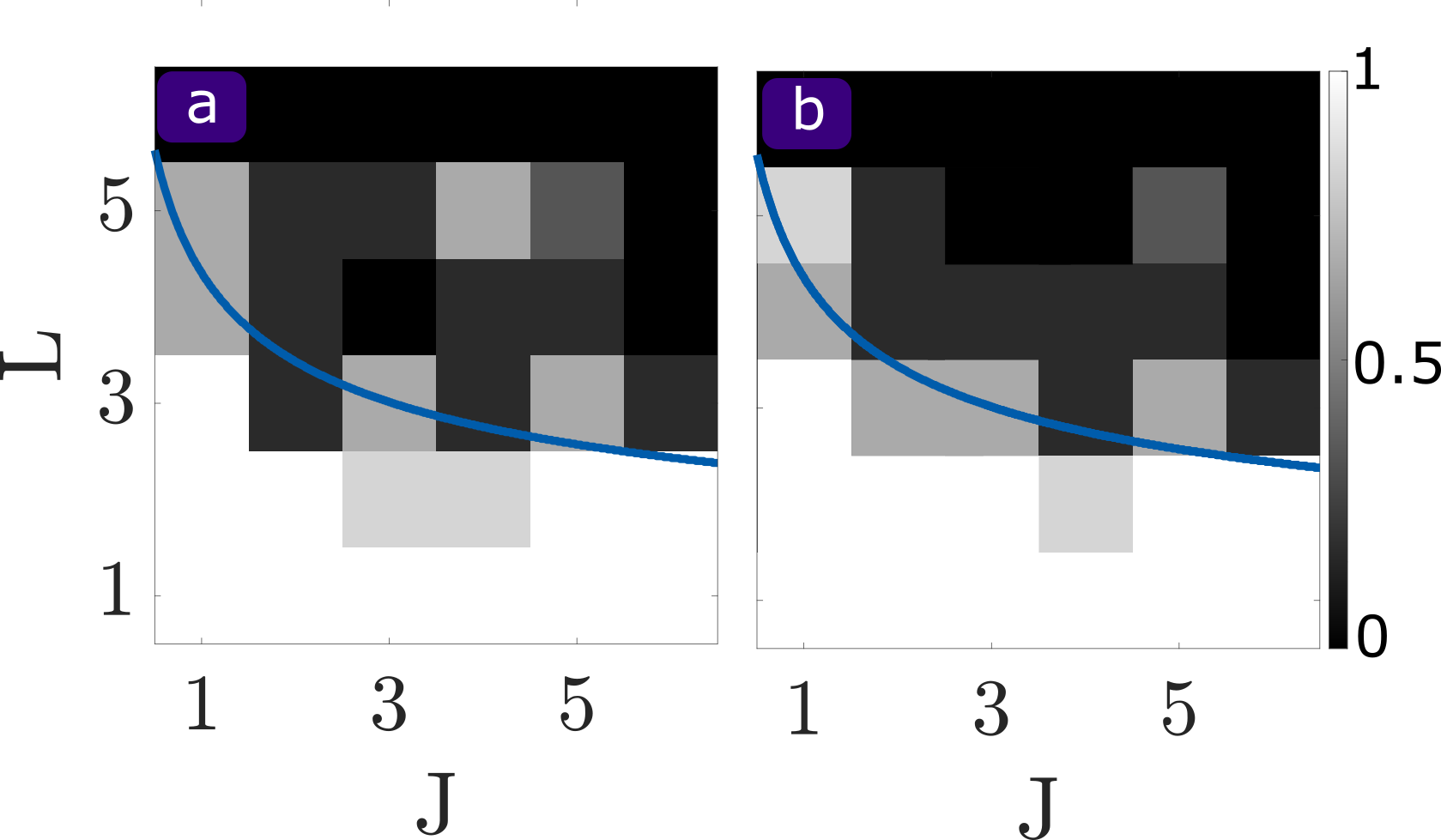}
    \caption{Success rate with respect to the number of targets/paths $L=Q$ and subspace dimension $J$ for (a) radar and (b) communications. The curve in blue shows the theoretical performance as predicted by Theorem \ref{th:main}.}
    \label{fig:JL}
\end{figure}

We evaluated our model and methods through numerical experiments using the CVX SDP3 \cite{grant2009cvx} solver for the problem in \eqref{dual_opt}. Throughout all experiments, we generated the columns of the transformation matrices $\mathbf{T}$ and $\mathbf{D}_p$ as 
$\mathbf{b}_n= [1, e^{\mathrm{j}2\pi \sigma_n}, \dots, e^{\mathrm{j}2\pi(J-1) \sigma_n}]$, where $\sigma_n \sim \mathcal{N}(0,1)$. The parameters $\bsym{\alpha}_r$ and $\bsym{\alpha}_c$ were drawn from a normal distribution with $|[\bsym{\alpha}_r]_\ell|=|[\bsym{\alpha}_c]_q|=1$, $\forall q$ and 
$\ell$. The real and imaginary components of the coefficient vectors $\mathbf{u}$ and $\mathbf{v}$ followed a uniform distribution over the interval $[0,1]$. 

We first evaluated our approach for specific channel realizations. 
\\
\noindent\textbf{Unequal number of targets and paths:}  Fig.~\ref{fig:results_dual} displays the radar and communications 3-D dual polynomials obtained for $M = 13$, $P=9, N_R = 5,L=4,Q=2$, and $J=3$. The parameter estimates are localized in the 3-D delay-Doppler-DoA hyperplane when the polynomials' $\ell_2$ norm achieves unity. We obtain perfect parameter estimation for both radar and communications channel parameters. 
\\ 
\noindent\textbf{Closely spaced targets:}  Fig.~\ref{fig:results_closely_parameters} depicts the recovery in a scenario when the delay-Doppler-DoA pairs of both channels are closely located in  $[0,1]^3$ with $M=17$, $P=11$, $N_R = 9$, and the subspace size $J=5$. Here, the minimum separations between the delay-times, Dopplers, and DoAs were set to $\Delta_\tau = \frac{0.1}{M}$, $\Delta_\nu = \frac{0.1}{P}$, and $\Delta_\beta = \frac{0.1}{N_R}$, respectively. The corresponding maximum separations were  $\frac{0.3}{M}$, $\frac{0.3}{P}$, and $\frac{0.3}{N_R}$.
$\bsym{\tau}_r = [0.67, 0.63]$, $\bsym{\nu}_r = [0.13,0.1]$, $\bsym{\beta}_r = [0.32,0.28]$, and $\bsym{\tau}_c = [0.53,0.49]$,  $\bsym{\nu}_c = [0.83, 0.8]$ , $\bsym{\beta}_c=[0.18,0.14]$. Fig. \ref{fig:results_closely_parameters} shows a near-perfect recovery of the channel parameters, where the small error is due to the closeness of the parameters. 

\begin{figure*}[!t]
    \centering
    \includegraphics[width = \textwidth]{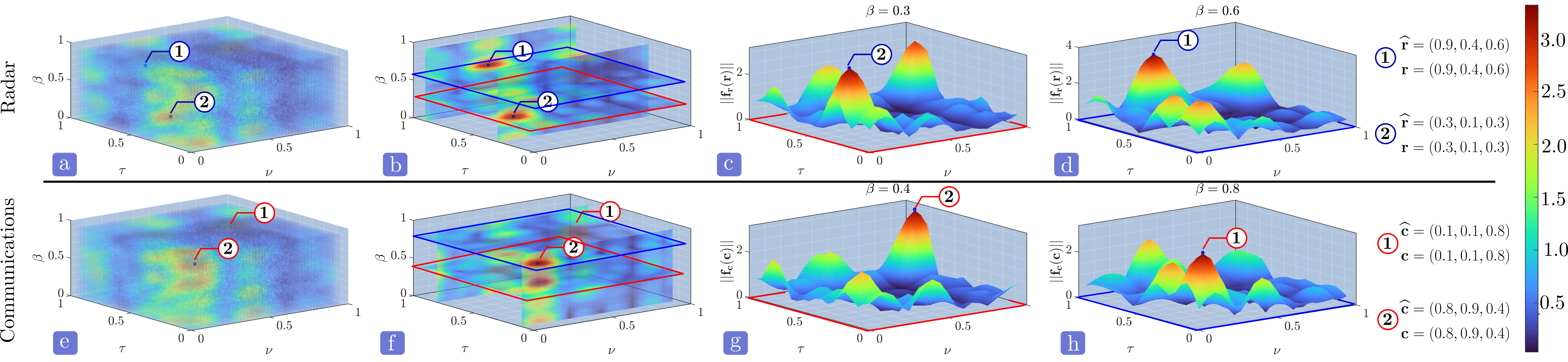}
    \caption{As in Fig. \ref{fig:results_dual}, but with steering vector errors and noise. Here, $L=Q=1$ targets/paths, $M=11$ samples, $P=7$ pulses/messages, subspace size $J=3$, SNR = 15 dB, and gain/phase error standard deviation $\sigma=0.01$.}
    \label{fig:pol_error}
\end{figure*}
\begin{figure}[!t]
    \centering
    \includegraphics[width =0.8 \linewidth]{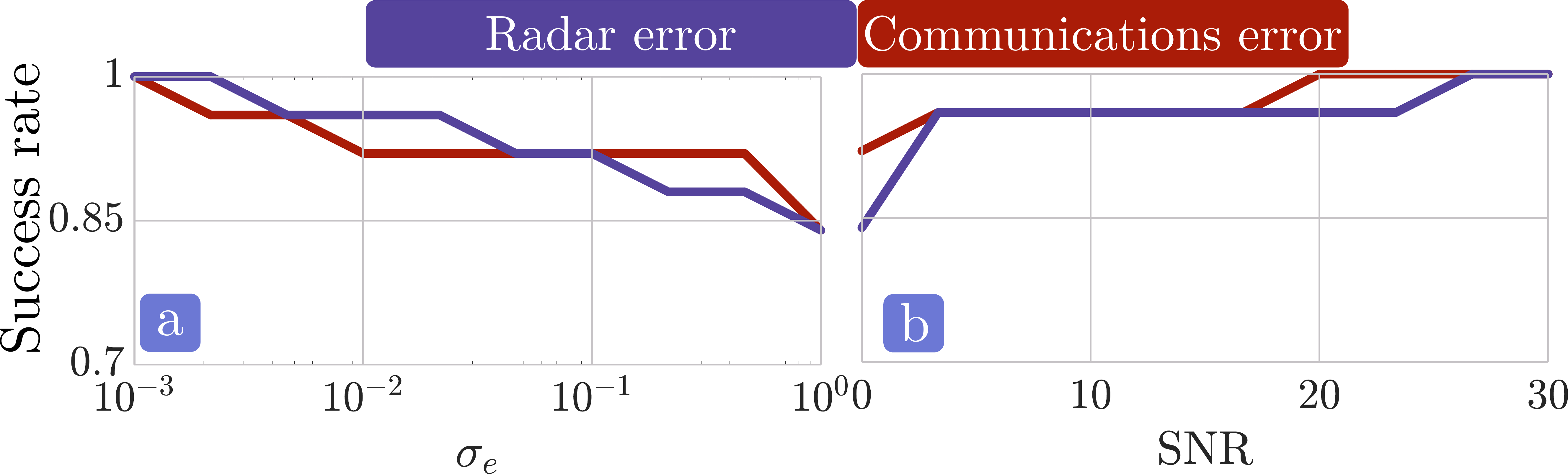}
    \caption{Success rate with respect to (a) standard deviations of the phase and gain errors in the steering vector (b) SNR. }
    \label{fig:stats_error}
\end{figure}
\noindent\textbf{Statistical performance:} We studied the statistical performance of the method by varying the number of samples $M$, the number of pulses/messages $P$, and receivers $N_R$ for $L=Q=2$. We define the radar error as  $\Vert\mathbf{X}_r-\widehat{\mathbf{X}}_r\Vert_F/\Vert \mathbf{X}_r\Vert_F$ and the communications error  $\Vert\mathbf{X}_c-\widehat{\mathbf{X}}_c\Vert_F/\Vert \mathbf{X}_c\Vert_F$. We declare success in the recovery when the radar and communications error is less than $10^{-3}$. Then, the success rate is the average of successful recovery over many trials. Fig. \ref{fig:stats}a, b, and show the success rate over 50 trials with respect to the number of samples $M$, the number of pulses $P$, and the number of received antennas, respectively. {\begin{remark}Note that the success rate of the communications estimation is better than radar for small values of $P$ and the contrary is true for large $P$. This is explained by the fact that the communications
signal transmits different waveforms at every PRI. Thus, increasing the number of pulses implies
estimating more transmitted messages. \end{remark} }

{Next, we varied the number of targets/paths $L=Q$ and the subspace size $J$ while fixing $M=P=N_R=7$. We computed the estimation error and the success recovery as in previous experiments. Figure \ref{fig:JL} shows the mean success rate over 50 trials. The recovery performance matches the overlaid theoretical curve derived from Theorem \ref{th:main}.}

\noindent\textbf{Noise and gain/phase errors:} 
To evaluate the effect of noise and steering vector errors in the general SDP of Section~\ref{sec:errors}, we generate the gain error, phase error, and noise as $\mathbf{n} \sim \mathcal{N}(0,\sigma^2) $, $\bsym{\phi} \sim \mathcal{N}(0,\sigma^2)$, and $\mathbf{w} \sim \mathcal{N}\left(0,\frac{\Vert \mathbf{y} \Vert_2^2}{10^{\frac{\textrm{SNR}}{10}}}\right)$, where SNR denotes the signal-to-noise ratio. The parameters $\mu_r$ and $\mu_c$ were chosen according to \eqref{eq:mu_r} and \eqref{eq:mu_c}. Fig. \ref{fig:pol_error} shows the radar and communications channel parameter localization in the 3-D dual polynomial for $L=Q=1, J=3, M=11, P=7, N_R=5$. Note that the maximum value of the polynomial's norm is achieved close to $\mu_r^2(1-(\varepsilon_e + 2\sqrt{MPN_R}) \epsilon_e\Vert \mathbf{Q}\Vert_2)$
 and $\mu_c^2(1-(\varepsilon_e + 2\sqrt{MPN_R}) \epsilon_e\Vert \mathbf{Q}\Vert_2)$ accordingly to proposition \ref{prop:dual_cert_errors} thereby demonstrating exact parameter recovery.  

 In Fig. \ref{fig:stats_error}, we analyze the performance for different $\sigma$ and SNRs. We set $L=Q=1, J=3, M=11, P=7, N_R=5$, using the same success criteria as in the previous experiment over 40 trials. The results suggest that the recovery performance does not decrease significantly for high variance in the steering vector errors as well as low SNRs. 
 
 \section{Summary and Discussion}
\label{sec:summ}
We solved the highly ill-posed multi-antenna DBD problem in the context of overlaid JRC application using the theories of PhTPs and SoMAN formulation. 
For simplicity, this formulation assumed synchronized radar and communications transmission. However, it is easily extended to a general case when one signal is delayed with respect to the other. For instance, when the radar transmission is delayed by $\tau_s$, the overlaid received signal becomes
\begin{align}
    [\mathbf{y}]_{\widetilde{m}} = &\suml[\bsym{\alpha}_r]_\ell [\mathbf{s}]_n e^{-\mathrm{j}2\pi(n([{\bsym{\tau}}_r]_\ell+\tau_s)+ p[{\bsym{\nu}}_r]_\ell +m[\bsym{\beta}_r]_\ell)}{+}\sumq [\bsym{\alpha}_c]_q[\mathbf{g}_p]_{n} e^{-\mathrm{j}2\pi(n[{\bsym{\tau}}_c]_q+ p[{\bsym{\nu}}_c]_q +m[\bsym{\beta}_c]_q)}.
    \label{eq:y_ps}
\end{align}
Note that this DBD problem is solved similarly to the steering vector error scenario of Section~\ref{sec:errors}, except that only the radar signal trail is affected. 

Further, our DBD formulation also holds when the number of radar pulses and communications messages are different (or the PRI $T_r$ is not equal to symbol duration $T_c$). As an example, assuming a common clock unit, one may consider the PRI to be an integer multiple of the symbol duration, i.e., $T_r = IT_c$ where $I\in \mathbb{N}_+$ this also implies 
$P_c = IP_r$. The communications transmit signal becomes 
\begin{equation}
x_c(t) = \sum_{i=1}^{I}\sum_{p=1}^{P_r} v_{i+pI}\left(t-(i+pI)\frac{T_r}{I} \right).
\end{equation}
Then, the SDP remains the same except that more communications symbols need to be recovered. This requires a larger subspace dimension than the radar signal and, hence, more samples are needed for perfect recovery. 

In the case of multiple radar and communications emitters, the received signal becomes 
$\mathbf{y} = \sum_{i=1}^{I_r}\mathcal{B}_{r
_i}(\mathbf{X}_{r_i}) +\sum_{i=1}^{I_c}\mathcal{B}_{c_i}(\mathbf{X}_{c_i}) $, where $I_r$ and $I_c$ are number of radar and communications sources, respectively. The primal optimization problem in \eqref{eq:primal_problem} becomes
\begin{align*}
    &\minimize_{\stackrel{\mathbf{X}_{r_1},\cdots, \mathbf{X}_{r_{I_r}}}{\mathbf{X}_{c_1}\cdots,\mathbf{X}_{c_{I_c}}}} \sum_{i=1}^{I_r} \Vert \mathbf{X}_{r_i}\Vert_{\mathcal{A}_r} +\sum_{i=1}^{N_c} \Vert \mathbf{X}_{c_i}\Vert_{\mathcal{A}_c} \text{ subject to } \mathbf{y} = \sum_{i=1}^{I_r}\mathcal{B}_{r
_i}(\mathbf{X}_{r_i}) +\sum_{i=1}^{I_c}\mathcal{B}_{c_i}(\mathbf{X}_{c_i}).
\end{align*}
The corresponding dual problem is 
\begin{align*}
&\maximize_{\mathbf{q}}\langle\mathbf{q,y}\rangle_{\mathbb{R}}\nonumber\\&\text{subject to } \Vert\mathcal{B}_{r_1}^\star(\mathbf{q})\Vert^\star_{\mathcal{A}_r}\leq1\;\cdots, \Vert \mathcal{B}_{r_{I_r}}^\star(\mathbf{q})\Vert^\star_{\mathcal{A}_r}\leq1\nonumber\\
    &\hphantom {\text {subject to } } \Vert \mathcal{B}_{c_1}^\star(\mathbf{q})\Vert^\star_{\mathcal{A}_c}\leq1, \cdots, \Vert \mathcal{B}_{c_{I_c}}^\star(\mathbf{q})\Vert^\star_{\mathcal{A}_c}\leq1.
\end{align*}
We identify that the $I_r+I_c$ constraints above employ, respectively, the PhTPs
$\mathbf{f}_{r_i}(\mathbf{r}) = \mathcal{B}_{r_{i}}^\star(\mathbf{q})\mathbf{w}_r(\mathbf{r})$, $i=1,\dots,I_r$, and $\mathbf{f}_{c_i}(\mathbf{c}) = \mathcal{B}_{c_{i}}^\star(\mathbf{q})\mathbf{w}_r(\mathbf{c})$, $i=1,\dots,I_c$.
The corresponding n-tuple SDP is obtained by replacing the constraints in the dual problem with $N_R+N_c$ LMIs as
\begin{align}
    &\maximize_{\mathbf{q,Q}}\quad \langle\mathbf{q,y}\rangle_{\mathbb{R}}\nonumber\\
    &\text{subject to }\mathbf{Q}\succeq 0\nonumber\\&\hphantom{\text{subject to }}  
    \begin{bmatrix}
        \mathbf{Q} & \widehat{\mathbf{Q}}_{r_1}^H \\
        \widehat{\mathbf{Q}}_{r_1} & \mathbf{I}_J 
        \end{bmatrix}
    \succeq0,\;\cdots,\; \begin{bmatrix}
        \mathbf{Q} & \widehat{\mathbf{Q}}_{r_{I_r}}^H \\
        \widehat{\mathbf{Q}}_{r_{I_r}} & \mathbf{I}_J 
        \end{bmatrix}
    \succeq0,\nonumber\\&\hphantom{\text{subject to }} 
    \begin{bmatrix}
        \mathbf{Q} & \widehat{\mathbf{Q}}_{c_1}^H \\
        \widehat{\mathbf{Q}}_{c_1} & \mathbf{I}_{JP} 
        \end{bmatrix}\succeq 0,\;\cdots,\;\begin{bmatrix}
        \mathbf{Q} & \widehat{\mathbf{Q}}_{c_{I_c}}^H \\
        \widehat{\mathbf{Q}}_{c_{I_c}} & \mathbf{I}_{JP} 
        \end{bmatrix}\succeq 0,
    \nonumber\\&\hphantom{\text{subject to }}
    \text{Tr}(\boldsymbol{\Theta}_\mathbf{n}\mathbf{Q}) = \boldsymbol{\delta}_{\mathbf{n}}.\label{m-dbd}
\end{align}
For numerical and theoretical results of this multiple emitter case, we refer the reader to \cite{10445951}. However, it remains the analysis under the presence of gain-phase errors for this challenging scenario. 

\bibliographystyle{elsarticle-num}
\bibliography{references}

\end{document}